\documentclass[aps,prb,twocolumn,showpacs,apsrev]{revtex4-1}
\usepackage[T1]{fontenc}
\usepackage{lmodern}
\usepackage[english]{babel}
\usepackage{amsmath}
\usepackage{amsfonts}
\usepackage{graphicx} 
\usepackage{comment}
\usepackage{caption}
\usepackage{subcaption}
\usepackage{xfrac}
\usepackage{color}
\usepackage{multirow}
\usepackage{hyperref}
\usepackage{braket}
\usepackage{xcolor}  
  
\hypersetup{
	colorlinks = true,
	pdftitle = {},
	pdfauthor = {},
	pdfkeywords = {},
	linkcolor = blue,
	citecolor = blue,
	filecolor = black,
	urlcolor = magenta
}

\newcommand{\smo}{SrMnO$_3$}
\newcommand{\sbmo}{Sr$_{1-x}$Ba$_{x}$MnO$_3$}
\newcommand{\sbmohalf}{Sr$_{0.5}$Ba$_{0.5}$MnO$_3$}

\graphicspath{{./figures/}}

\begin{document}

\title{Magnetic and ferroelectric properties of Sr$_{1-x}$Ba$_{x}$MnO$_3$ from first principles}
 
\author{Alexander Edstr\"om}\altaffiliation[]{Current address: Institut  de  Ci\`{e}ncia  de  Materials  de  Barcelona  (ICMAB-CSIC),  Campus  UAB,  08193  Bellaterra,  Spain \\ aleeds@kth.se}
\author{Claude Ederer}
\affiliation{Materials Theory, ETH Z\"urich, Wolfgang-Pauli-Strasse 27, 8093 Z\"urich, Switzerland}


\begin{abstract}

Density functional theory (DFT) calculations are used to study the magnetic and ferroelectric properties of \sbmo, with focus on $x=0.5$, under isotropic volume expansion/compression and biaxial strain. 
Our results indicate that, unexpectedly, Ba substitution alters the electronic structure in a way that, at fixed lattice parameter, notably enhances the interatomic magnetic exchange interactions.
However, increasing Ba-content also causes a volume expansion which tends to weaken 
these interactions, leading to a net effect of weakly suppressed magnetism, as observed in experiments. 
The ferroelectric properties, on the other hand, are found to be less affected by changes in the electronic structure and can largely be understood in terms of the volume expansion caused by Ba-substitution. The calculated electric polarization as a function of biaxial strain in \sbmo{} for $x=0$ and $x=0.5$, shows that the difference between the two is mainly due to differences in the magnetic order at certain strain values, accompanied by enormous magnetoelectric coupling. 

\end{abstract}

\maketitle

\section{Introduction}\label{sec.intro}

After first principles calculations suggested that either strain or Ba doping can turn the otherwise G-type antiferromagnetic (AFM) paraelectric material SrMnO$_3$ ferroelectric (FE)~\cite{PhysRevLett.104.207204,PhysRevB.79.205119}, and thus also multiferroic, the material has been the subject of significant research activities.  
Experimental studies have confirmed that biaxial tensile strain leads to ferroelectricity~\cite{Becher2015,acs.nanolett.5b04455,PhysRevB.97.235135,PhysRevLett.107.137601} and the interest in the material is further boosted by predictions of a ferromagnetic (FM) ferroelectric (FE) phase at large strain~\cite{PhysRevLett.104.207204} and expectations of pronounced magnetoelectric coupling, related to strong spin-phonon coupling~\cite{PhysRevB.84.104440,PhysRevB.85.054417,PhysRevLett.109.107601,Sakai_et_al:2012,PhysRevB.89.064308,Goian_et_al:2017}.
Recent computational work investigated the complete strain and temperature-dependent ferroic phase diagram of perovskite-structured SrMnO$_3$ and showed the existence of a tetracritical point, with coinciding magnetic and ferroelectric ordering temperatures at a certain biaxial tensile strain, where novel magnetoelectric coupling effects are expected~\cite{PhysRevMaterials.2.104409}. This recently culminated in the prediction of an enormous magnetoelectric coupling and interesting new caloric effects around this point~\cite{2019arXiv190512955E}. While BaMnO$_3$ is not stable in the cubic perovskite structure, Sr$_{1-x}$Ba$_x$MnO$_3$ has been synthesized with $x$ up to around 0.5 and several experimental~\cite{PhysRevB.90.140401,doi:10.1021/acsami.5b06478,ADMI201601040,doi:10.1063/1.5090824} and computational~\cite{PhysRevLett.109.107601,PhysRevB.90.220405,PhysRevB.94.165106,PhysRevMaterials.2.084404} studies of multiferroicity in this compound have appeared. Furthermore, a combination of strain engineering and chemical substitution has been used to tune the properties of the material. 

So far, Ba-substitution has largely been discussed as means of expanding the lattice, i.e., causing a chemically induced strain. The assumption has thus been that the $A$-site cation does not notably influence the chemical bonding and electronic structure around the Fermi energy. In particular, recent experimental work~\cite{doi:10.1063/1.5090824} studied the magnetic ordering temperature of \sbmo~ thin films as function of both $x$ and strain. They observed that, in the regime where the material remains G-type antiferromagnetic (AFM), the magnetic ordering temperature decreases monotonically with increasing unit cell volume in a similar manner, regardless whether this volume expansion is caused by chemical substitution or epitaxial strain. This observation supports the idea that the effect of strain or Ba substitution is very similar in \sbmo, even though these two ways of tuning the materials properties should differ in at least two ways: i) strain results in a structural symmetry lowering avoided by chemical substitution, and ii) chemical substitution can also cause changes in the electronic structure and, furthermore, introduces effects related to substitutional disorder.  

In this work, we use density functional theory (DFT) calculations to further scrutinize the possible similarities and differences between chemical substitution and strain on the magnetic and FE properties of \sbmo. 
The results indicate that Ba substitution, somewhat unexpectedly, alters the electronic structure in a way which enhances Mn-O hybridization, and consequently enhances magnetic exchange interactions. The FE properties are comparatively less affected by the chemical influence. However, at certain values of biaxial strain dependent on $x$, the magnetic order changes, which markedly affects the electric polarization via the strong magnetoelectric coupling. 

The paper is structured as follows. In Sec.~\ref{sec.meth} we describe the computational methods used in this work. We then present our results in Sec.~\ref{sec.results}, starting with calculations of magnetic and FE properties of SrMnO$_3$ (SMO), Sr$_{0.5}$Ba$_{0.5}$MnO$_3$ (SBMO) and BaMnO$_3$ (BMO) as function of isotropic volume expansion in Sec.~\ref{sec.vol}, thereby focusing on magnetic exchange interactions and phonon instabilities. Next, we concentrate on the effect of biaxial strain on the magnetic and ferroelectric properties of SBMO in Sec.~\ref{result.strain} and compare these results to previous calculations for SMO~\cite{PhysRevMaterials.2.104409}. Finally, we present a concluding discussion in Sec.~\ref{sec.concl}.

\section{Computational Methods}\label{sec.meth}

Density functional theory (DFT) calculations are performed using the projector augmented wave (PAW) method~\cite{PhysRevB.50.17953,PhysRevB.59.1758} implemented in the \emph{Vienna ab-initio simulation package} (VASP)~\cite{KRESSE199615,PhysRevB.49.14251,PhysRevB.47.558}. All settings are chosen similar to previous calculations for SMO~\cite{PhysRevMaterials.2.104409}. The plane wave energy cut-off is set to 680 eV and a mesh of at least $7 \times 7 \times 7$ $\mathbf{k}$-points is used for the basic perovskite unit cell, or correspondingly for larger supercells. 
Sr$_{0.5}$Ba$_{0.5}$MnO$_3$ (SBMO) is described by a supercell containing two perovskite units, with Sr and Ba atoms arranged in a three-dimensional checkerboard pattern, similar as in previous computational studies of this material~\cite{PhysRevB.94.165106,PhysRevMaterials.2.084404}. 

We use the PBEsol version of the generalised gradient approximation~\cite{PhysRevLett.100.136406} (GGA) as exchange-correlation functional, with an additional Coloumb repulsion~\cite{PhysRevB.57.1505} of $U_\mathrm{eff}=3~\mathrm{eV}$ on the Mn $d$-electrons.
The combination of PBEsol and $U_\mathrm{eff}=3~\mathrm{eV}$ (or a slightly smaller value) has been suggested for $A$MnO$_3$, $A$=Ca, Sr, Ba, in Ref.~\onlinecite{PhysRevB.85.054417}, based on comparison of the energy difference between the FM and G-type AFM states with those from hybrid functional calculations, and has then been used in a number of studies of these materials~\cite{Becher2015,marthinsen_2016,PhysRevMaterials.2.104409,2019arXiv190512955E,SMO_Jij}. 
A recent study~\cite{PhysRevB.93.205110} also argued that PBEsol is more suitable than the spin polarized PBE functional~\cite{PhysRevLett.77.3865} or local density approximation (LDA) for describing \sbmo, although this study also suggest that other combinations, with non-spin-polarized exchange-correlation functionals can be preferable for certain properties.

Magnetic exchange coupling parameters $J_{ij}$ are calculated by mapping the DFT total energies from different magnetic configurations on a Heisenberg Hamiltonian:
\begin{equation}\label{eq.Heis}
H = - \sum_{i<j} J_{ij} \hat{m}_i \cdot \hat{m}_j \quad ,
\end{equation}
where $\hat{m}_i$ specifies the direction of the magnetic moment on Mn site $i$. The exchange coupling parameters are then obtained similarly as in previous studies~\cite{PhysRevB.84.224429,PhysRevMaterials.2.104409}:
\begin{equation}
  \label{eq.Jij}
J_{ij} = -\frac{E_{\uparrow \uparrow} + E_{\downarrow \downarrow} - E_{\uparrow \downarrow} - E_{\downarrow \uparrow}}{4n} \quad ,  
\end{equation}
where $E_{\sigma_i \sigma_j}$ denotes the total energy of having spin $i$ in spin state $\sigma_i$  and spin $j$ in state $\sigma_j$, while all other spins are fixed in a given reference state, here kept as G-type AFM for consistency. The integer $n$ is a multiplicity factor due to the limited size of the unit cell. For these calculations we use a $2 \times 2 \times 2$ supercell of the basic perovskite cell. Recently, a comprehensive study of the magnetic exchange interactions in \smo, and comparison between Eq.~\eqref{eq.Jij} and another method for calculating $J_{ij}$ was presented in Ref.~\onlinecite{SMO_Jij}.

We consider biaxial tensile strain by fixing the lattice parameters in the ``in-plane'' $x$ and $y$ directions to a value $a$, while the ``out-of-plane'' lattice parameter, $c$, along the $z$ direction, is allowed to relax. FE structures are calculated by shifting the atomic positions slightly according to the unstable phonon mode at the $\Gamma$ point (in the centrosymmetric structure) and subsequently relaxing the atomic coordinates as well. The resulting electric polarization is then calculated using the Berry phase formalism~\cite{PhysRevB.47.1651}. Atomic coordinates are relaxed until all forces on the atoms are below 10$^{-4}$ eV/\AA. Phonon frequencies are obtained within the harmonic approximation using a finite difference approach, i.e., by calculating the forces when individual atoms are shifted by 0.015~\AA. Phonons were calculated at the $\Gamma$ point of a doubled unit cell corresponding to rhombohedral lattice vectors, in order to accommodate the G-type antiferromagnetism and the checkerboard arrangement of Ba/Sr cations. This corresponds to phonons at the $\Gamma$ and R points of the basic cubic perovskite Brillouin zone. 

\section{Results}\label{sec.results}

\subsection{Cubic volume expansion}\label{sec.vol}

As outlined in the introduction, it is often assumed that SMO and BMO have a very similar electronic structure around the Fermi energy, since the Sr$^{2+}$ and Ba$^{2+}$ cations do not contribute any valence states in that energy region, which instead is dominated by Mn $d$ and O $p$ states. Hence, one might expect that the main effect of substituting Ba into SMO is a lattice expansion, due to the larger ionic radius of Ba$^{2+}$ compared to Sr$^{2+}$. 

To test the validity of this assumption, we begin our study by comparing SMO, SBMO and BMO within the cubic perovskite structures.
The calculated equilibrium lattice parameters are $a_\mathrm{SMO}=3.79~\AA$, $a_\mathrm{SBMO} = 3.84~\AA$ and $a_\mathrm{BMO} = 3.89~\AA$. For cubic Sr$_{1-x}$Ba$_{x}$MnO$_3$, the lattice constant has been experimentally measured to increase from 3.807~\AA~to 3.856~\AA~at 300~K~\cite{PhysRevLett.107.137601}, as $x$ increases from 0 to 0.4.~\footnote{As mentioned in Sec.~\ref{sec.intro}, the synthesis of \sbmo~ in the cubic perovskite structure is only possible up to $x \approx 0.5$. We nevertheless include the case of perovskite-structured BMO here, analogous to Ref.~\onlinecite{PhysRevB.79.205119}, to allow for a more systematic comparison.} This indicates rather good agreement with our calculated values, which however, are slightly underestimated 
using PBEsol+$U_\text{eff}=3$\,eV 
(assuming that the thermal expansion is small for temperatures between 0 and 300 K). The increase in the lattice constant with increasing $x$ is also slightly underestimated in the calculations. Comparison between the total energies obtained for different magnetic orders (including ferromagnetic as well as G, C and A-type AFM) indicates G-type AFM as lowest energy state for each compound (i.e., all nearest neighbor Mn spins are antiferromagnetically aligned~\cite{Wollan/Koehler:1955}). 

\begin{figure}
	\centering
	\includegraphics[width=0.5\textwidth]{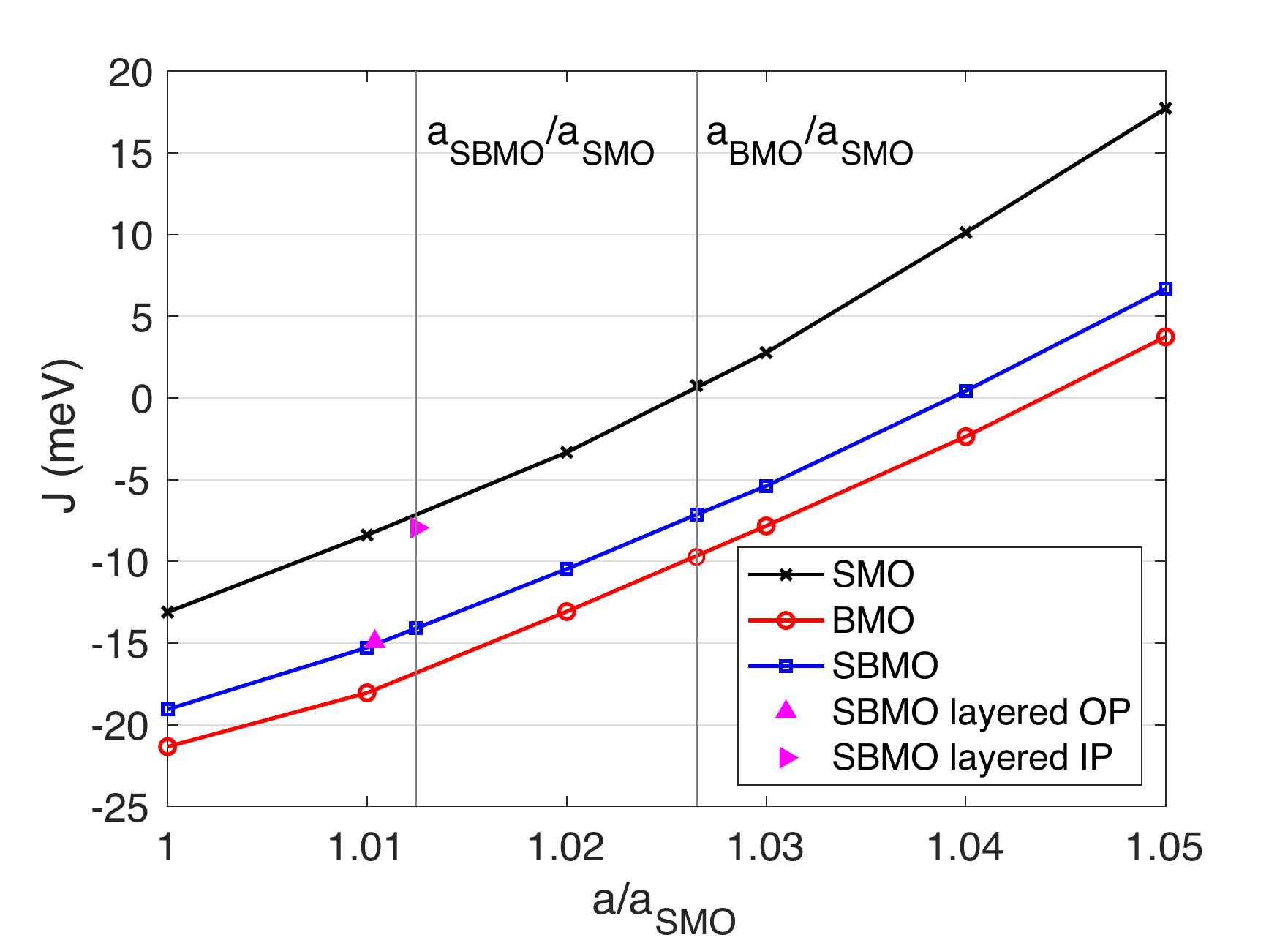}
	\caption{Nearest neighbor exchange interaction $J_1$ for SMO (black), SBMO (blue), and BMO (red) under isotropic volume expansion, plotted as function of the cubic lattice parameter $a$ relative to the equilibrium lattice constant for SMO, $a_\text{SMO}$. The calculated equilibrium lattice parameters for SBMO and BMO are indicated by thin vertical lines. For SBMO, results are also shown for the IP and OP NN exchange interactions in a (fully relaxed) layered structure. }
	\label{fig.J_of_vol}
\end{figure}

Next, we calculate the magnetic nearest neighbour (NN) exchange interactions in cubic SMO, SBMO, and BMO as functions of lattice parameter. Results are shown in Fig.~\ref{fig.J_of_vol}. The interaction parameter $J_1$ is negative at each of the equilibrium lattice parameters for each compound, leading to the G-type antiferromagnetism. Furthermore, in each compound, $J_1$ shows the same qualitative trend with volume expansion, first decreasing in magnitude and eventually changing sign, resulting in the stabilization of a ferromagnetic state at large volumes. This behavior of $J_1$ as function of interatomic distance has been discussed previously for SMO, both under epitaxial strain~\cite{PhysRevMaterials.2.104409} and isotropic volume expansion~\cite{SMO_Jij}. In Ref.~\onlinecite{SMO_Jij}, the mechanism behind this trend was analyzed in detail and found to be related to a lowering in energy of the Mn $e_\mathrm{g}$ states, as the crystal field splitting is reduced for larger lattice parameters. This leads to enhanced hybridization between Mn $e_\mathrm{g}$ and O $p$ states, yielding a positive contribution to the exchange interaction~\cite{SMO_Jij}.

While the qualitative trends for $J_1$ as function of lattice parameter, seen in Fig.~\ref{fig.J_of_vol}, are the same for all three compositions, there is also a clear quantitative shift towards more negative values with increasing Ba substitution (when compared at the same lattice parameter). This causes a significantly stronger (negative) exchange interaction at small lattice parameters, and a corresponding increase in the value of the lattice parameter where the transition to positive $J_1$ occurs. Thus, the effect of Ba substitution on the magnetic properties is two-fold. First, it increases the lattice parameter, which weakens the negative (AFM) magnetic exchange coupling. Second, it affects the electronic structure and chemical bonding in a way such that it enhances this coupling at a given lattice parameter. As these two effects partially cancel out, the total effect of substituting Sr with Ba, on the magnetic exchange interaction, is relatively small. 

At the respective equilibrium lattice parameters, $J_1$ is slightly larger in magnitude for SBMO than for SMO, while for BMO  it is smaller than for SMO. 
In mean field theory, the magnetic ordering temperature $T_\mathrm{c}$ is proportional to the magnetic exchange interactions. Thus, considering only the NN interaction, one would expect $T_\mathrm{c}$ to first increase and then decrease with Ba-substitution, with the $T_\mathrm{c}$ of BMO being lower than that of SMO. 
The increase in $T_\mathrm{c}$, when going from SMO to SBMO, indicated by our calculations might seem inconsistent with the experimental observation of a weak monotonous decrease in the magnetic ordering temperature of bulk \sbmo~ with increasing Ba content~\cite{PhysRevLett.107.137601}. However, the experimental results also show that, while the lattice parameter increases linearly with $x$, $T_\mathrm{c}$ is rather unaffected by Ba substitution for small $x$, and only decreases notably for $x \geq 0.2$. This supports the idea that there is a competition between the effect of the lattice expansion and a further chemical influence on the magnetic exchange interactions, leading to a rather small net effect.

Apart from uncertainties stemming from the slight underestimation of lattice parameters in our calculations, or from a weak dependence of the magnetic exchange constants on the magnetic reference state used to compute them (see, e.g., Ref.~\onlinecite{SMO_Jij}), another source of uncertainty, which could affect the balance between the two competing mechanisms, stems from the insufficient description of chemical disorder. 
To investigate this further, Fig.~\ref{fig.J_of_vol} also contains $J_1$ calculated for Sr$_{0.5}$Ba$_{0.5}$MnO$_3$ with a (fully relaxed) layered arrangement of Sr and Ba instead of a checkerboard pattern. This results in a tetragonal symmetry with different lattice parameters perpendicular (IP) and parallel (OP) to the tetragonal axis.
The IP $J_1$ in this case clearly deviates from the results of SBMO in the checkerboard pattern (blue line, squares), which can be taken as an indication that a more sophisticated description of the order/disorder of Sr and Ba atoms might indeed affect the results. Further investigations in this direction might therefore be of interest. 

\begin{figure}
	\centering
	\includegraphics[width=0.5\textwidth]{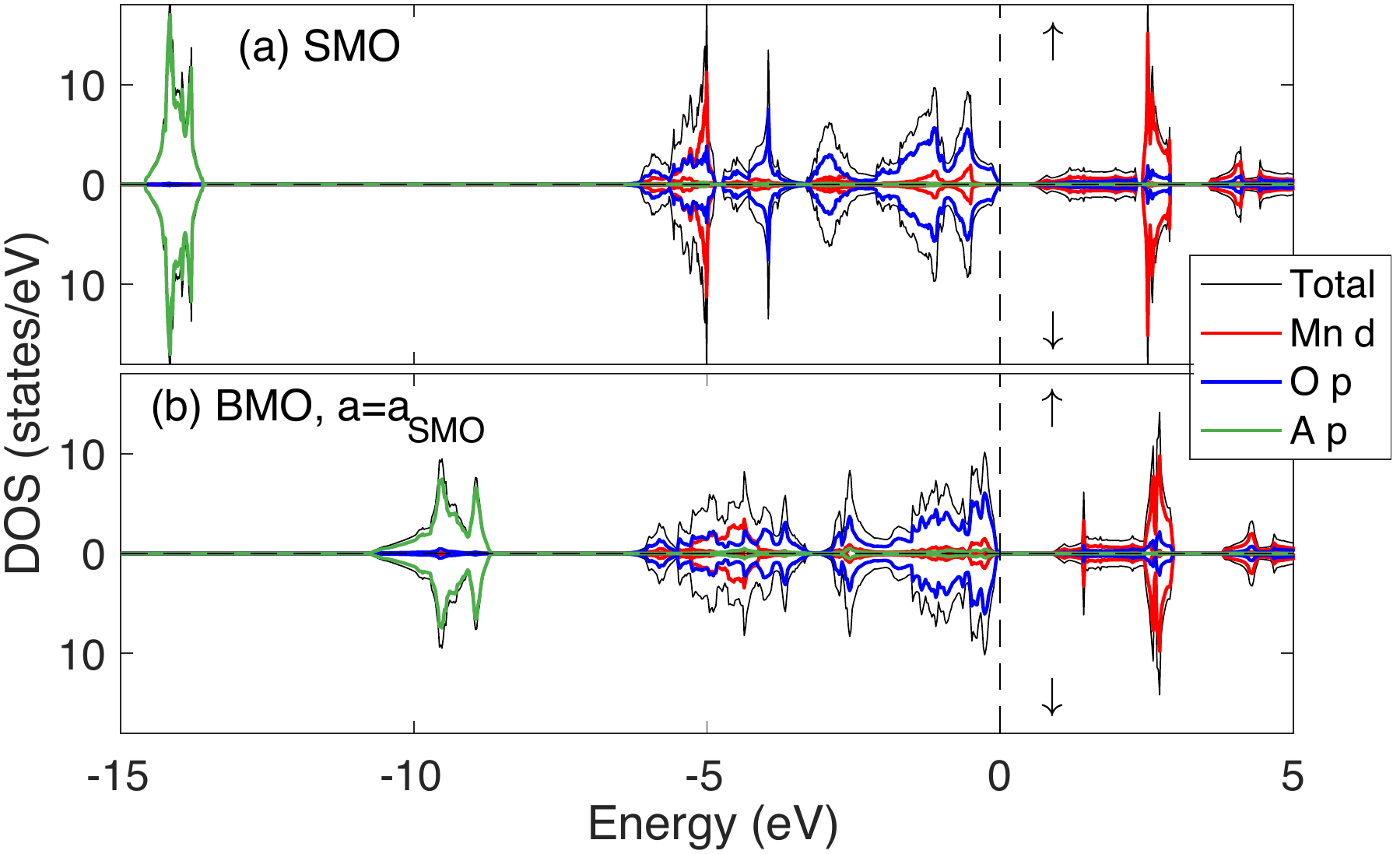}
	\caption{Spin polarized total (black), Mn $d$ (red), O $p$ (blue) and $A$-site $p$ (green) densities of states (DOS), relative to the top of the valence band, for cubic SMO ($A$=Sr) in (a) and BMO ($A$=Ba) in (b), both calculated at the SMO equilibrium lattice parameter.}
	\label{fig.dos}
\end{figure}

Nevertheless, since, from Fig.~\ref{fig.J_of_vol}, Ba-substitution is seen to notably enhance the magnetic exchange coupling at fixed lattice parameter, it must have an effect on the electronic structure beyond that caused by the simple lattice expansion. To analyze this further, Figs.~\ref{fig.dos}(a) and (b) show the densities of states (DOS) for SMO and BMO calculated at the same lattice constant, $a_\mathrm{SMO}$, respectively. As expected, the DOS in the two cases are similar, with the top of the valence band and bottom of the conduction band dominated by O $p$ and Mn $d$ states. However, important differences can be identified. For SMO [Fig.~\ref{fig.dos}(a)], one can identify a rather sharp, localized Mn $d$ peak at approximately $-5$\,eV, which corresponds to the occupied local majority spin $t_{2g}$ states. For BMO [Fig.~\ref{fig.dos}(b)] this peak is much broader, indicating a stronger delocalization of the Mn $d$ states, resulting in an enhanced hybridization with the O $p$ states. In a superexchange picture, where the magnetic exchange interaction is mediated by hopping between Mn $d$ and O $p$ states, this would indeed be expected to enhance the exchange interaction. 

At lower energy, one finds the $A$-site $p$ semi-core states. One can see that the Ba $5p$ states are higher in energy compared to to the Sr $4p$ states, as expected for electronic states with higher principal quantum number. Furthermore, the semi-core $p$ states of Ba are broadened compared to the Sr $p$ states. 
To further analyze the role of these $A$-site semi-core $p$ states on the strength of the magnetic coupling in SMO and BMO, we construct maximally localized Wannier functions~\cite{Marzari_et_al:2012} for the whole energy region shown in Fig.~\ref{fig.dos}, and also including the O $s$ states at around $-17$\,eV. The resulting quadratic spread of the Ba $p$-like Wannier functions is about 50\,\% larger than for the Sr $p$ states. In both cases the leading  matrix elements of the Wannier Hamiltonian connecting the $A$-site $p$ states to the $s$ and $p$ orbitals of the surrounding oxygen atoms are non-negligible and are significantly larger for BMO (up to 0.74\,eV compared to a maximum of 0.57\,eV in SMO). 
On the other hand, applying an empirical potential shift to the Ba $p$ states and shifting them down in energy to approximately $-14$\,eV, i.e., to the same energy as the Sr $p$ states in SMO, does not have a noticeable effect on the energy difference between the FM and the G-type AFM state.
We thus conclude that the main effect is the larger spatial extension of the Ba $p$ states compared to the Sr $p$ states, which leads to stronger hybridization with the O levels and thus indirectly affects the hybridization between O $p$ and Mn $d$ states determining the magnetic coupling strength.

Based on the data presented in Fig.~\ref{fig.J_of_vol} and Fig.~\ref{fig.dos}, it is clear that the magnetic properties of Sr$_{1-x}$Ba$_x$MnO$_3$ cannot be understood only in terms of the volume expansion with increasing $x$. Instead, the Ba substitution also alters the electronic structure near the Fermi energy, thereby affecting (indirectly) the strength of the Mn-O hybridization even for fixed lattice constant. However, Ba substitution will not only affect the magnetic properties, but will also induce ferroelectricity in the material. To analyze this, we now calculate phonon frequencies as functions of isotropic volume expansion in the cubic perovskite structure. 

\begin{figure}
	\centering
	\includegraphics[width=0.5\textwidth]{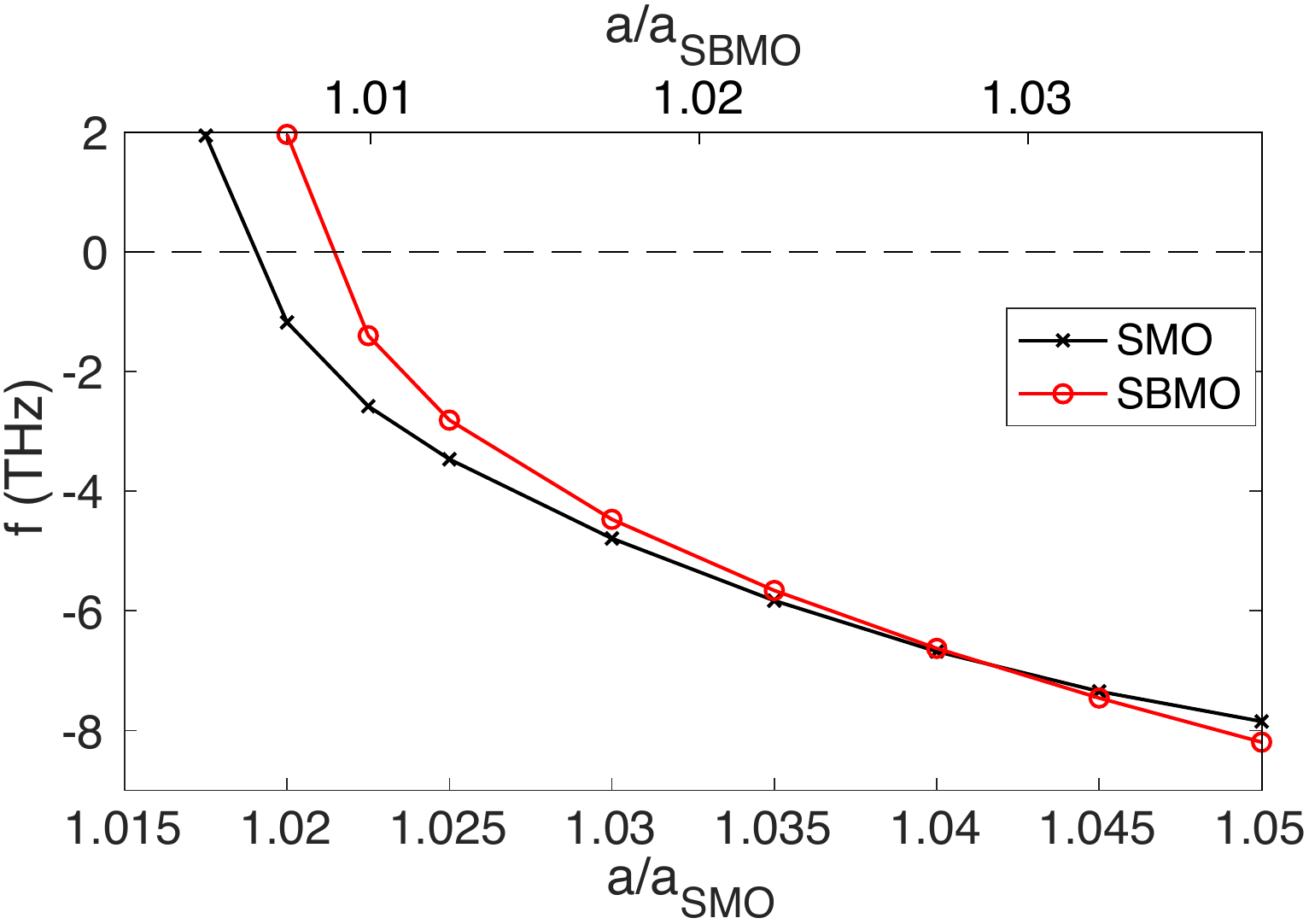}
	\caption{Frequency of lowest polar phonon mode in cubic SMO and SBMO as function of lattice parameter. Imaginary frequencies are plotted on the negative axis. }
	\label{fig.f_strain}
\end{figure}

Fig.~\ref{fig.f_strain} shows the lowest-lying polar phonon frequency at the $\Gamma$ point in SMO and SBMO, as function of isotropic strain. Both materials develop a FE instability, indicated by an imaginary phonon frequency, near 2\% strain (relative to $a_\text{SMO}$). Even though the FE instability in SMO appears at a slightly smaller lattice parameter than for SBMO, the dependence of the imaginary phonon frequency on the lattice parameter is very similar for both compounds above 2.5\,\% strain. This indicates that the ferroelectricity can be understood as being largely due to the volume expansion caused by the Ba substitution, in contrast to the magnetic properties where changes in the electronic structure also play a considerable role. This is consistent with results in Ref.~\onlinecite{Sakai_et_al:2012}, showing that the experimentally measured phonon frequencies of Sr$_{1-x}$Ba$_x$MnO$_3$ can be reproduced using first principles calculations, modelling the effect of Ba-substitution by changing the lattice parameter of SrMnO$_3$.

We note that, according to Fig.~\ref{fig.f_strain}, we find a non-polar ground state for 
the unstrained bulk of \sbmohalf, while experimentally it is known that \sbmo~is FE at $x=0.5$~\cite{PhysRevLett.107.137601}. This can again be explained by the slight underestimation of volume in PBEsol (plus $U_\text{eff}=3$\,eV). In contrast, recent work reported that PBE+$U$ calculations yield a polar ground state structure in \sbmohalf~\cite{PhysRevMaterials.2.084404}, which is consistent with the larger lattice parameter obtained within PBE. Ref.~\onlinecite{PhysRevMaterials.2.084404} also claims that 
PBEsol with various $U_\mathrm{eff}$ gives a non-polar ground state for \sbmohalf, even over a broad range of strain, which, however, appears inconsistent with our results shown in Fig.~\ref{fig.f_strain} and also our results for biaxial strain presented in Sec.~\ref{result.strain}.
Furthermore, in Ref.~\onlinecite{marthinsen_2016} it was reported that oxygen vacancies interact favorably with strain and Ba-doping, as well as FE polarization. Since oxygen vacancies are not considered in our work, nor most other computational studies, this is likely another factor contributing to discrepancy between experimental and computational results.
Thus, in order to be consistent with our previous calculations for \smo, we continue to use PBEsol with $U_\mathrm{eff}=3~\mathrm{eV}$, even though it appears to slightly overestimate the amount of Ba-substitution needed to turn \sbmo~FE, similarly as it has previously been seen to likely overestimate the critical strain at which \smo~turns FE~\cite{PhysRevMaterials.2.104409}. We still expect it to correctly describe the qualitative trends in magnetic and FE properties with strain and Ba-substitution, which is what we are interested in.

\subsection{Biaxial (epitaxial) strain}\label{result.strain}

\subsubsection{Energetics of FE and magnetic order}

\begin{figure}
	\centering
	\includegraphics[width=0.5\textwidth]{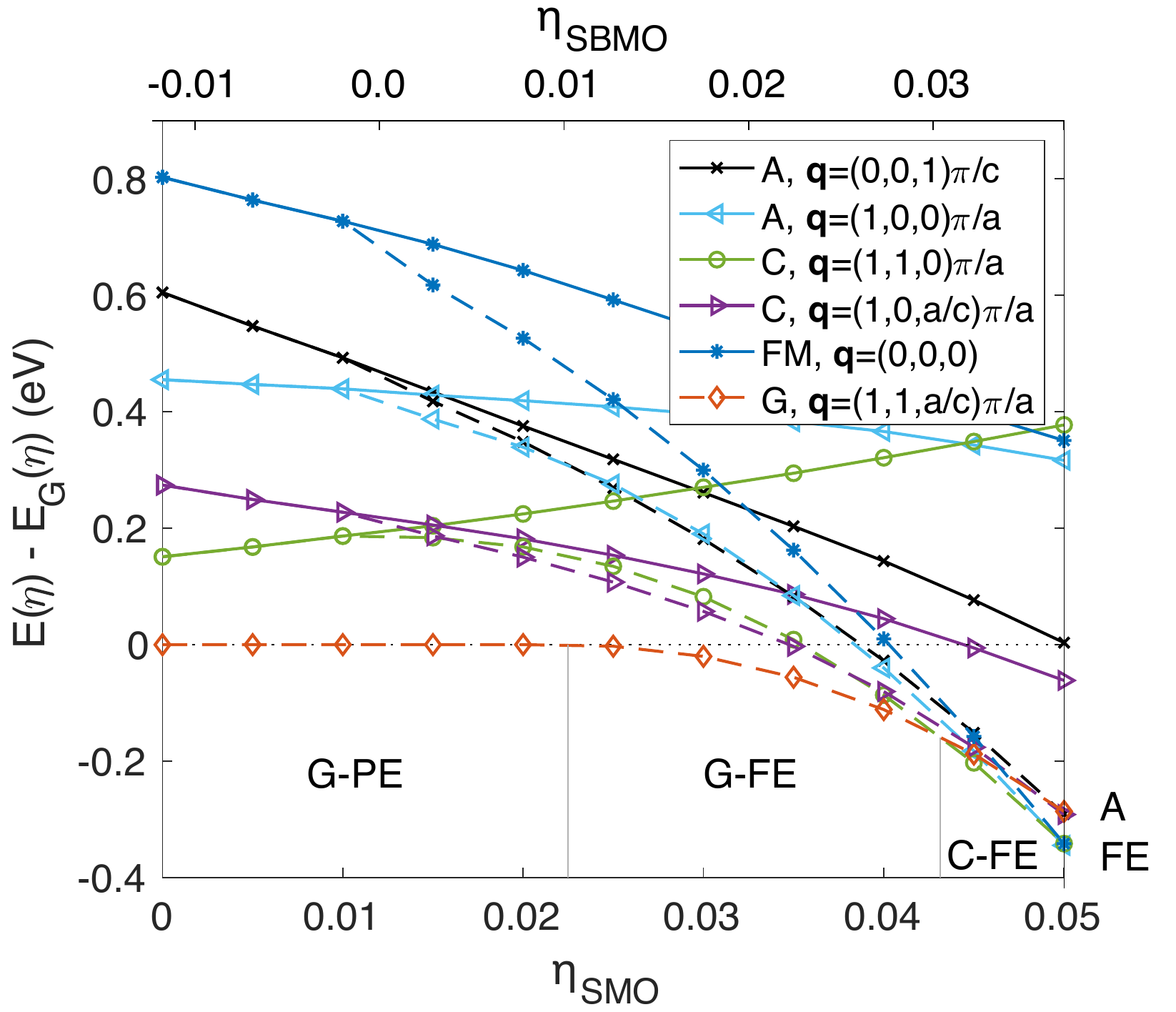}
	\caption{Total energy (per 40 atom supercell) of SBMO for different magnetic orders, relative to that of G-type AFM, as function of epitaxial strain with centrosymmetric (solid lines) or FE (dashed lines) structures. }
	\label{fig.Eofa}
\end{figure}

After analyzing the case of an isotropic volume expansion, we next study the effect of epitaxial strain on SBMO.
In order to better understand the interplay between such biaxial strain and Ba-doping on the magnetic and FE properties of \sbmo, we compare to corresponding data for SMO from Ref.~\onlinecite{PhysRevMaterials.2.104409}.  
Fig.~\ref{fig.Eofa} shows the calculated total energies of different magnetic configurations, including G, C and A-type AFM as well as FM, as function of biaxial tensile strain, defined either with respect to the SMO or SBMO equilibrium lattice parameters, $\eta_\mathrm{SMO} = a/a_\mathrm{SMO} - 1$ or $\eta_\mathrm{SBMO} = a/a_\mathrm{SBMO} - 1$, respectively, where $a$ is the in-plane (IP) lattice parameter. 
We consider two sets of calculations: i) keeping a centrosymmetric structure (solid lines) and ii) allowing non-centrosymmetric FE structural distortions (dashed lines). Analogous data for SMO has been presented and discussed in Ref.~\onlinecite{PhysRevMaterials.2.104409}. 

As mentioned in Ref.~\onlinecite{PhysRevMaterials.2.104409}, A and C-type AFM order each correspond to three degenerate $\mathbf{q}$-vectors in the case of a cubic crystal structure, that is $\mathbf{q}=(1,0,0)$, $(0,1,0)$ and $(0,0,1)$ for A and $\mathbf{q}=(1,1,0)$, $(1,0,1)$ and $(0,1,1)$ for C, with the $\mathbf{q}$-vectors given in units of $\pi/a$. Biaxial tensile strain lifts this degeneracy by making in-plane and out-of-plane directions inequivalent, while ferroelectric displacement can further lower the symmetry so that all three $\mathbf{q}$-vectors can become inequivalent. However, it is sufficient to consider one of the two degenerate $\mathbf{q}$-vectors and let the in-plane FE structure relax into the direction of polarization that yields the lowest energy for that magnetic $\mathbf{q}$-vector. 
Hence, we consider two A and two C-type magnetic orders, namely (0,0,1) and (1,0,0) for A and (1,1,0) and (1,0,1) for C.

As seen in Fig.~\ref{fig.Eofa}, G-AFM order is lowest in energy for small strain values, consistent with the cubic case discussed in Sec.~\ref{sec.vol}. SBMO turns FE just above $\eta_\mathrm{SMO}=2\%$ or $\eta_\mathrm{SBMO} \approx 1\%$, after which it remains FE at larger strains. For larger strains, the magnetic order becomes C$_{110}$ (subscript indicating the $\mathbf{q}$-vector of the magnetic order) at $\eta_\mathrm{SMO}$ just above 4\%, or $\eta_\mathrm{SBMO}\approx 3\%$, and then A$_{100}$ as $\eta_\mathrm{SMO}$ approaches $5\%$, where, however, the A$_{100}$, C$_{110}$ and FM magnetic orders are very near each other in energy. This behaviour can be compared to that of SMO, presented in Ref.~\onlinecite{PhysRevMaterials.2.104409}. SMO turns FE at a similar IP lattice parameter as SBMO, but changes from G to C$_{101}$ just above 3\% strain, and then becomes FM just over 4\% strain. Interestingly, the C-AFM order that appears in the intermediate strain region is different in SMO and SBMO. To the best of our knowledge, other studies of this material, except Ref.~\onlinecite{PhysRevMaterials.2.104409}, have not discussed the different possible types of A and C-type magnetic order in the strained and FE cases. 

In the regions with G or C$_{110}$ AFM order, the polarization is in the (110)-direction. However, at larger strain, where the compound turns A$_{100}$, the symmetry breaking of the magnetic order, via magnetostructural coupling, rotates the polarization by about 2.6$^\circ$ towards the (100) direction. The polarization as function of strain is discussed in more detail in Sec.~\ref{result.strain_P}. 

\subsubsection{Magnetic Exchange Interactions}\label{result.strain_J}

\begin{figure}
	\centering
	\includegraphics[width=0.5\textwidth]{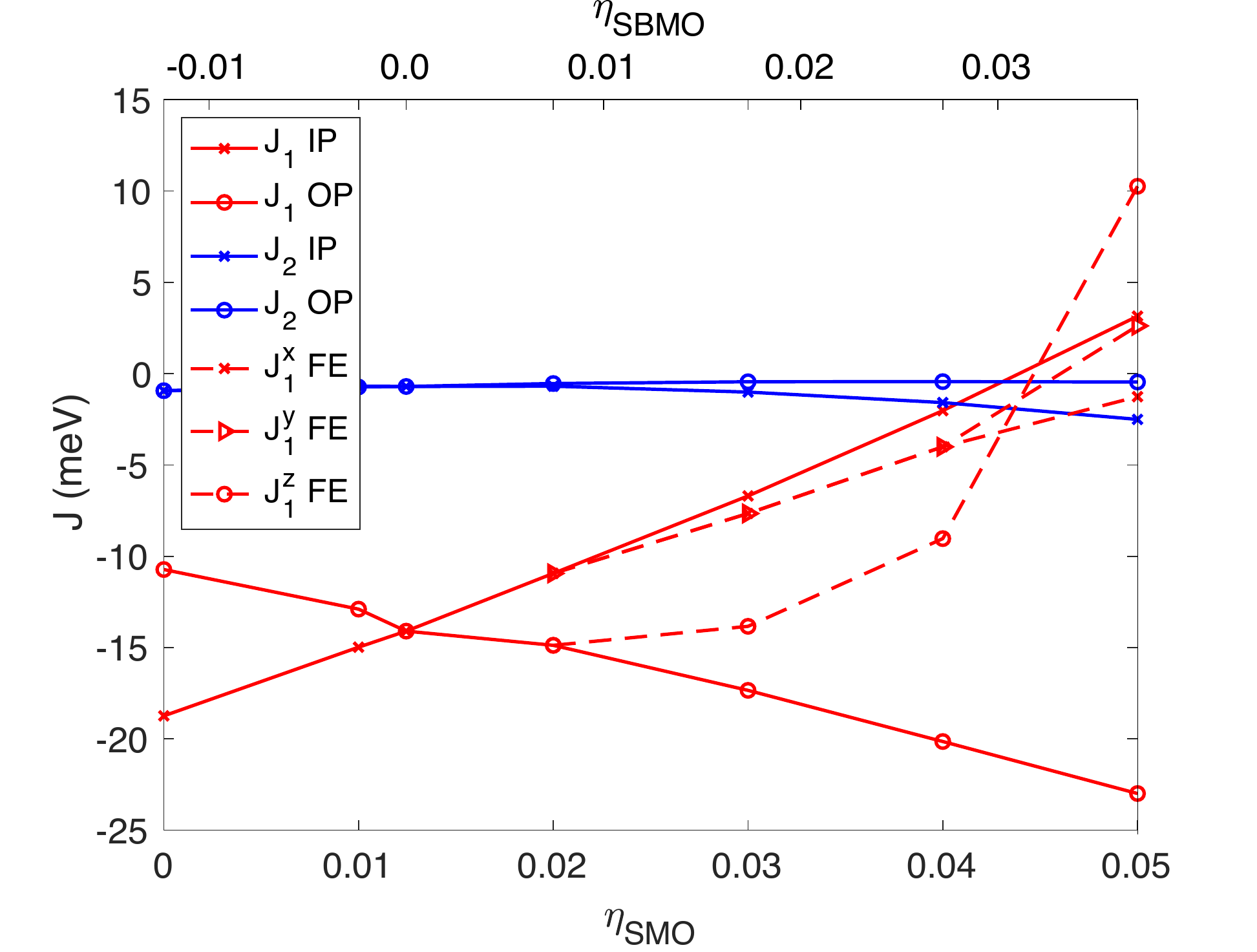}
	\caption{Nearest ($J_1$) and second nearest ($J_2$) neighbor exchange interactions of SBMO as functions of biaxial strain. With strain these split up into in-plane (IP) and out-of-plane (OP) interactions. Solid lines show the results for centrosymmetric tetragonal structures, while the dashed lines show the results for FE structures, where they are shown for all three bond directions $x$, $y$ and $z$.}
	\label{fig.Jij_strain}
\end{figure}

To understand the strain-induced changes in the magnetic order observed in Fig.~\ref{fig.Eofa}, we now calculate magnetic exchange interactions as function of strain, similar to what has been done for SMO in Ref.~\onlinecite{PhysRevMaterials.2.104409}. Fig.~\ref{fig.Jij_strain} shows the NN (red) and second NN (blue) exchange interactions, $J_1$ and $J_2$, for SBMO as functions of biaxial strain.
The IP NN exchange interaction, $J_1^\mathrm{IP}$, decreases in magnitude with increasing IP lattice parameter, and then changes sign for $\eta_\mathrm{SMO}$ just above 4\% ($a_\mathrm{SBMO} \approx 3\%$). In contrast, the out-of-plane (OP) NN exchange interaction increases in magnitude, as the OP lattice parameter contracts with increasing IP lattice parameter. From this, one would expect A or C-type AFM order at $\eta_\mathrm{SMO}=5\%$, depending on the competition between $J_1$ and $J_2$. However, for $\eta_\mathrm{SMO}>2\%$, a FE polar distortion is favored, and the NN exchange interactions calculated for these non-centrosymmetric structures are plotted with dashed lines. The effect of the FE distortion, with polarization in the plane, is small on $J_1^\mathrm{IP}$. However, the change in the OP Mn-O-Mn bond angle causes a drastic change in $J_1^\mathrm{OP}$, which now also changes sign just above $\eta_\mathrm{SMO}=4\%$. 
Hence, for $4\% \leq \eta_\mathrm{SMO} \leq 5\%$, the different first and second NN exchange interactions are all of similar size. Their competition leads to the C$_{110}$ state for $\eta_\mathrm{SMO}=4.5$\,\%, and then the A$_{100}$ state at $\eta_\mathrm{SMO}=5$\,\%, nearly degenerate with the C$_{110}$ and FM states. 

In Fig.~\ref{fig.J_of_vol} it was seen that $J_1$ of SMO and SBMO show the same qualitative trend with volume expansion, although there is quantitative shift between the two cases. A similar situation is now observed under biaxial strain, where the qualitative changes for SBMO are similar to those observed for SMO in Ref.~\onlinecite{PhysRevMaterials.2.104409}. 
In both SMO and SBMO, biaxial tensile strain causes $J_1^\mathrm{IP}$ to change sign because of the increase in the IP lattice parameter, whereas $J_1^\mathrm{OP}$ changes sign because of the change in Mn-O-Mn bond angle resulting from the FE distortion. We have seen that, at fixed lattice parameter, Ba substitution strengthens the initial magnitude of $J_1$, such that a larger IP lattice parameter is needed to cause the sign change in this exchange interaction for SBMO compared to SMO. On the other hand, the FE transition, which strongly affects $J_1^\mathrm{OP}$, occurs at similar IP lattice parameter for the two compounds. As a result, $J_1^\mathrm{IP}$ and $J_1^\mathrm{OP}$ change sign at almost the same $a$ in SBMO, whereas in SMO $J_1^\mathrm{IP}$ changes sign at a smaller $a$ than $J_1^\mathrm{OP}$. This leads to the differences in the magnetic order predicted for SBMO and SMO in the region $3\% \leq \eta_\mathrm{SMO} \leq 5\%$.

As SBMO turns FE with increasing strain, the polarization is along the [110]-direction, as long as it remains in the G or C$_{110}$ magnetic phases. However, as mentioned earlier, at $\eta_\mathrm{SMO}=5\%$, the IP symmetry breaking of the A$_{100}$ magnetic order causes a small rotation of the polarization, by $2.6^\circ$, towards the $x$-direction. This brings the Mn-O-Mn bond angle along the $x$-direction slightly closer to $180^\circ$, while the opposite occurs along the $y$-direction, causing the NN exchange interaction in the $x$-direction to be slightly more AFM than that in the $y$-direction.

\subsubsection{Volume and strain dependence of $T_c$}

As mentioned in Sec.~\ref{sec.intro}, Ref.~\onlinecite{doi:10.1063/1.5090824} presented a systematic experimental study of the magnetic ordering temperature of Sr$_{1-x}$Ba$_x$MnO$_3$ thin films as function of both $x$ and biaxial strain. 
It was found that the magnetic transition temperature decreases monotonously with increasing unit cell volume, regardless of whether the variation of the unit cell volume is due to strain or chemical substitution.
Based on this, it was suggested that, as long as the magnetic order remains G-type AFM, the unit cell volume is the key parameter determining the magnetic transition temperature.  
However, in light of the significantly stronger $J_1$ obtained for SBMO compared to SMO at the same lattice constant (Fig.~\ref{fig.J_of_vol}), and the large difference between $J_1^\text{IP}$ and $J_2^\text{OP}$ due to symmetry-breaking biaxial strain (Fig.~\ref{fig.Jij_strain}), a simple dependence of the magnetic ordering temperature on unit cell volume does not seem to be consistent with our computational results.

Hence, in order to compare our results more closely to the observations made in Ref.~\onlinecite{doi:10.1063/1.5090824}, we now analyze the effect of epitaxial strain on the magnetic ordering temperature in both SMO and SBMO.
We determine the magnetic ordering temperature within mean field theory using the first and second NN exchange interactions plotted in Fig.~\ref{fig.Jij_strain} for SBMO and those obtained in Ref.~\onlinecite{PhysRevMaterials.2.104409} for SMO. 

\begin{figure}
	\centering
	\includegraphics[width=0.5\textwidth]{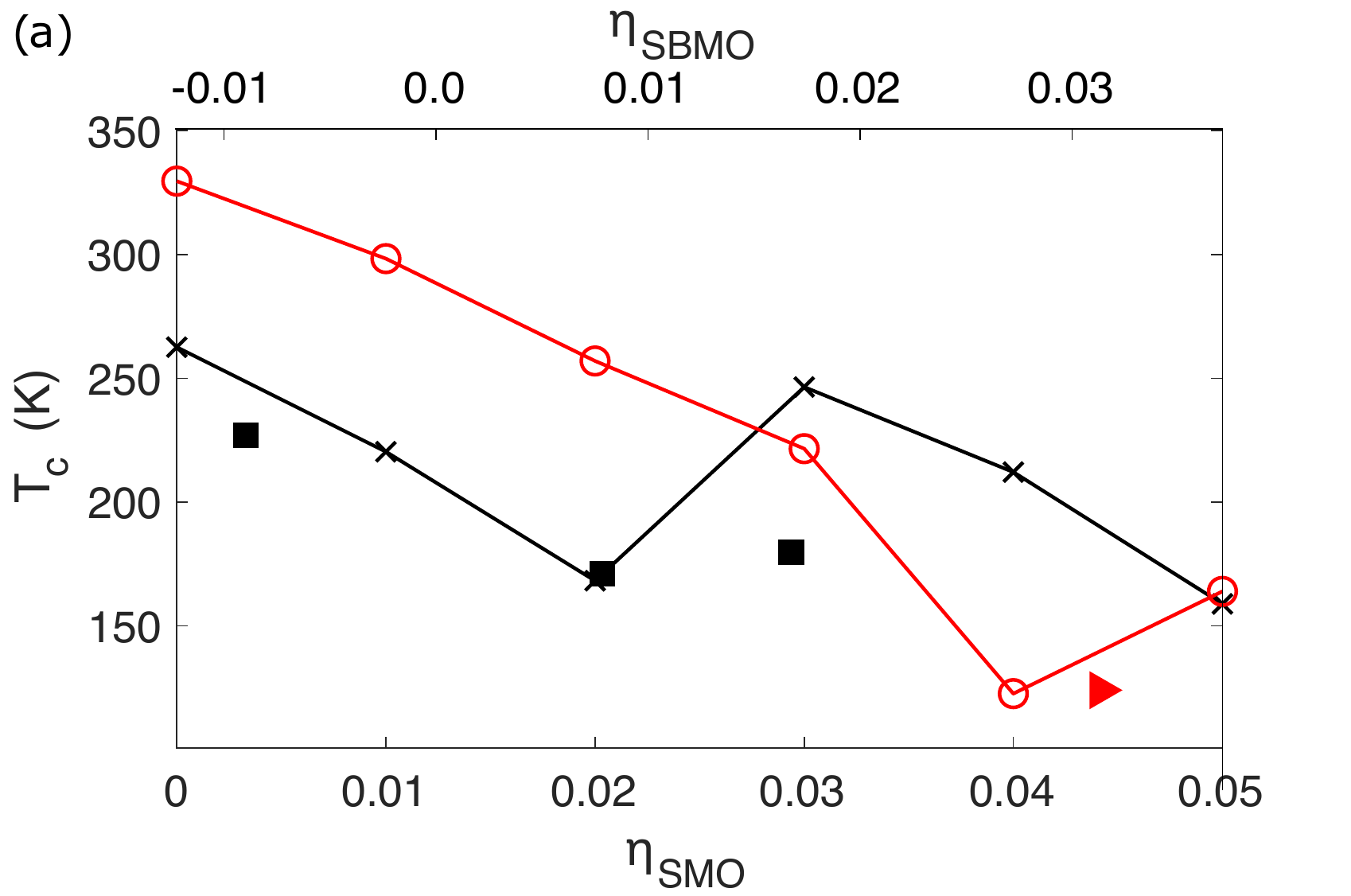}
    	\includegraphics[width=0.5\textwidth]{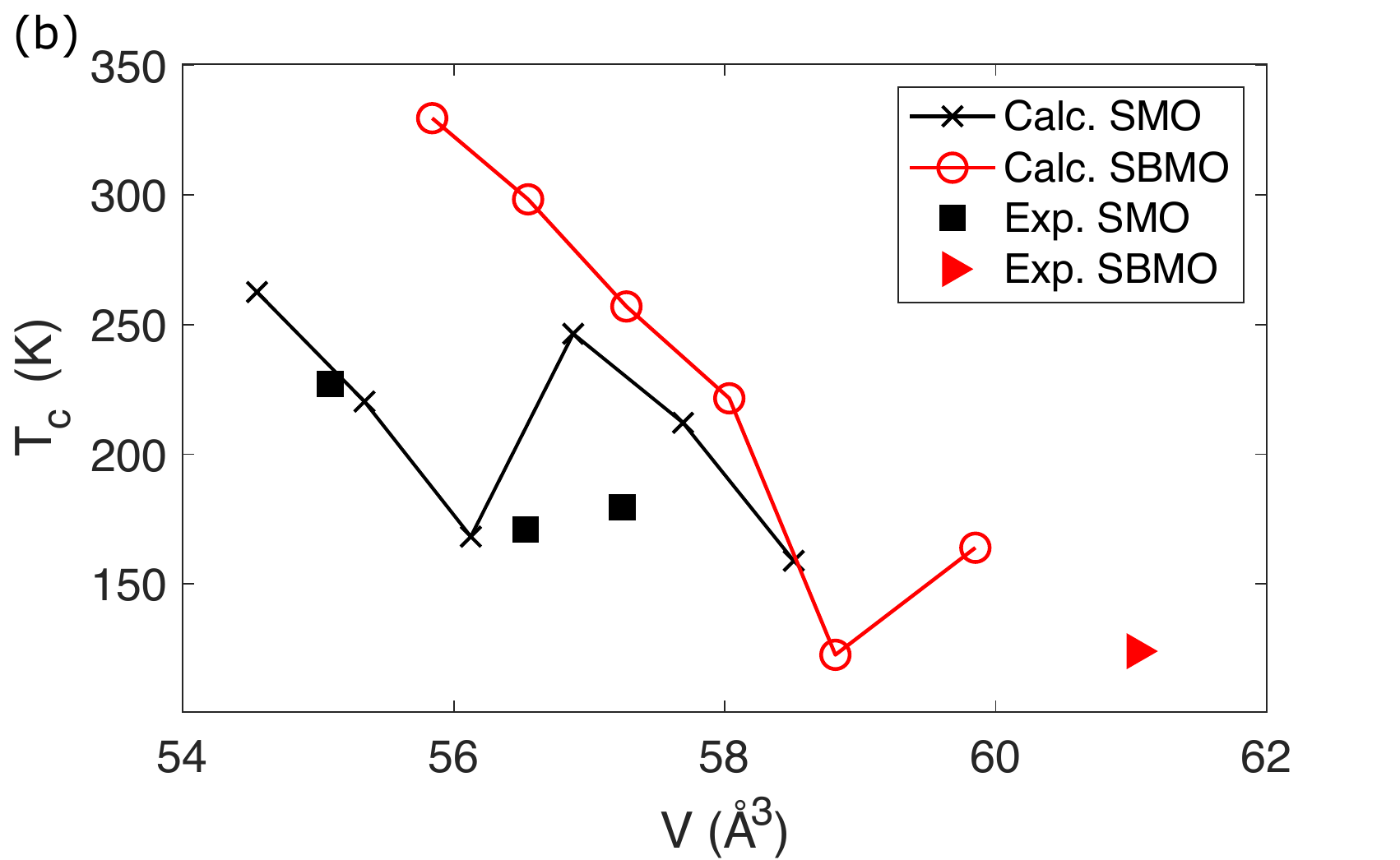}
	\caption{(a) Calculated magnetic ordering temperature, obtained from mean field theory, as function of biaxial tensile strain in SMO and SBMO, together with experimental data from Ref.~\onlinecite{PhysRevB.92.024419,doi:10.1063/1.5090824}. (b) Magnetic ordering temperatures as function of unit cell volume, varying due to biaxial tensile strain, in SMO and SBMO.}
	\label{fig.Tc_strain}
\end{figure}

Fig.~\ref{fig.Tc_strain}(a) shows the calculated magnetic ordering temperatures as functions of biaxial tensile strain, for both SMO (black) and SBMO (red). Experimental data is also shown for comparison~\cite{doi:10.1063/1.5090824,PhysRevB.92.024419}. 
In both SMO and SBMO, $T_\mathrm{c}$ decreases with increasing IP lattice constant, in the strain region where the magnetic order remains G-AFM. 
In that region, $J_1^\mathrm{IP}$ decreases in magnitude but remains negative, while the magnitude of $J_1^\mathrm{OP}$ increases due to the contraction along $c$. Due to the larger number of IP bonds, their effect dominates and thus $T_\mathrm{c}$ decreases.
At certain strains, the magnetic ordering temperatures in both SMO and SBMO jump to a higher values. 
For SMO, this occurs when the system becomes FE, but remains G-AFM. For higher strains, $T_\mathrm{c}$ decreases again, even when the magnetic order changes.
In contrast, for the case of SBMO, $T_\mathrm{c}$ continues to decrease with increasing strain as the material turns FE above $\eta_\text{SMO}=2$\,\%. Instead, the upturn in $T_\text{c}$ occurs at a larger IP lattice constant, where the magnetic order changes 

Plotting the same data as function of unit cell volume rather than in-plane expitaxial strain [see Fig.~\ref{fig.Tc_strain}(b)], shows qualitatively similar behavior, since the volume increases monotonously with strain, as the decrease in the OP lattice parameter $c$ is insufficient to compensate the increase in the IP lattice constant $a$.
Most notably, the magnetic ordering temperatures calculated for SMO and SBMO do not coincide in the region where their volumes overlap. This is consistent with the enhancement of the magnetic NN interaction due to Ba substitution already found for the cubic case (see Fig.~\ref{fig.J_of_vol}). 

To conclude, our calculated data reproduces the decrease in $T_\mathrm{c}$ with increasing strain/volume in SMO in the region with G-AFM order, as well as the upturn where the magnetic order changes to C-type. However, it also indicates that the magnetic ordering temperature in Sr$_{1-x}$Ba$_{x}$MnO$_3$ does not only depend on unit cell volume and magnetic order, but is also strongly affected by changes in the electronic structure caused by varying $x$. 
We note that the experimental data does not contain data points for different $x$ but similar volume, and thus there is in fact no clear contradiction between our results and the available experimental data. Further experimental characterization, including more combinations of strain and $x$, and perhaps further computations for intermediate values of $x$ would be required to resolve this in more detail.

\subsubsection{Electric Polarization}\label{result.strain_P}

From Figs.~\ref{fig.J_of_vol}-\ref{fig.f_strain}, it follows that Ba-substitution has a notable chemical influence on the magnetic exchange interactions, whereas the ferroelectricity is less affected by this chemical influence and might largely be understood in terms of the change in volume. To further investigate this, we next compare the ferroelectric properties of SMO and SBMO under biaxial tensile strain. 

Two sets of calculations are performed for each compound. One where the structural relaxations at each strain are performed with G-type AFM order and another one where the structure is relaxed using the magnetic order that yields the lowest energy at the corresponding strain value. In both cases, OP lattice parameter as well as atomic positions are relaxed, the latter initialized with a small off-centering, in the [110]-direction, to allow the system to develop a FE polarization. 
The electric polarization for both relaxed structures is then calculated using G-type AFM magnetic order, to ensure an insulating gap in the electronic structure, thus neglecting the effect of the different magnetic orders on the electronic contribution to the Born effective charges.

\begin{figure}
	\centering
	\includegraphics[width=0.5\textwidth]{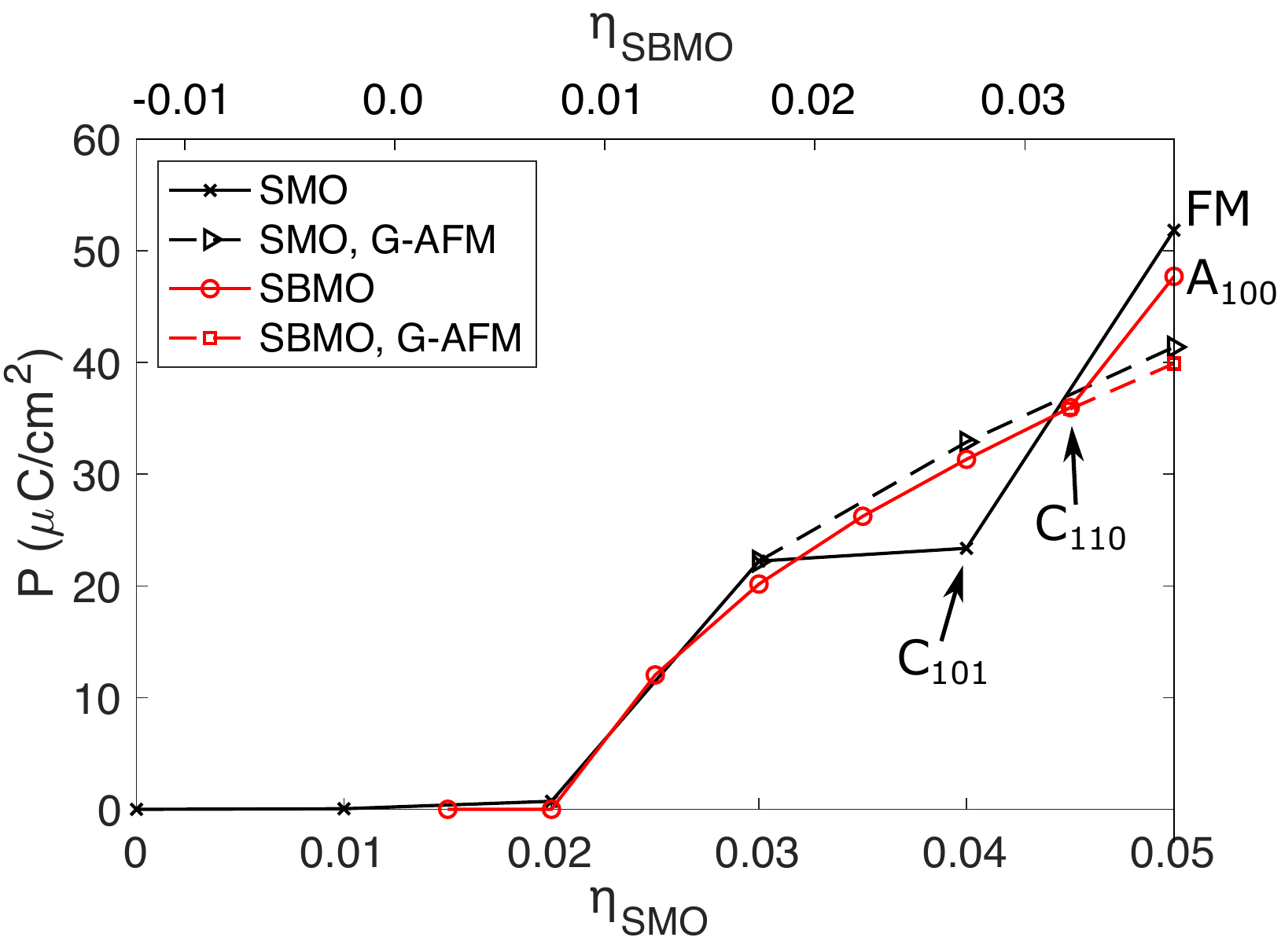}
	\caption{Polarization as function of biaxial tensile strain in SMO (black) and SBMO (red). Results are shown for structures that were optimized with the magnetic order yielding the lowest energy at a given strain (solid lines, crosses for SMO and circles for SBMO). For large strains, where the lowest energy structure is no longer G-AFM, results are also shown for structures that were relaxed with G-AFM (dashed lines, triangles for SMO and squares for SBMO).}
	\label{fig.P_strain}
\end{figure}

Fig.~\ref{fig.P_strain} shows the FE polarization of SMO (black) and SBMO (red) as function of IP lattice parameter.
For the structures relaxed with G-AFM order (dashed lines), the polarization of SMO and SBMO is very similar at all strains. This indicates that, for a fixed lattice parameter in the direction of polarization, the polarization is indeed barely affected by chemical substitution. 
On the other hand, it can be seen that for the cases where the magnetic order deviates from G-AFM (solid lines, $\eta_\text{SMO} \geq 4$\,\% for SMO and $\eta_\text{SMO} \geq 4.5$\,\% for SBMO), the resulting effect on the structure can lead to  drastic changes in the FE polarization. This indicates the strong magnetoelectric coupling in this system. A strong variation in the polarization of SBMO depending on magnetic order was recently also reported based on hybrid functional calculations in  Ref.~\onlinecite{okuyama2020}.
As discussed recently in Ref.~\onlinecite{2019arXiv190512955E}, the magnetoelectric coupling in SMO depends strongly on both strain and magnetic order. Furthermore, at large strain and low temperatures, the polarization is large and higher order coupling terms become important. 
At $\eta_{SMO}=4\%$, where SMO becomes C$_{101}$-AFM, the polarization of SMO is significantly reduced compared to that with G-AFM order, indicating that the magnetoelectric coupling at that strain is such that C$_{101}$ order and polarization disfavor each other, at least compared to the G-AFM order~\cite{higherLambdaNote}. 
In contrast, the SBMO polarization with C$_{110}$ order at $\eta_\text{SMO}=4.5\%$ is nearly the same as that with G-AFM order. At $\eta_\text{SMO}=5\%$, where SMO and SBMO are FM and A$_{100}$ ordered, respectively, both have strongly enhanced polarization compared to the G-type AFM case, indicating that these magnetic orders couple favorably to the electric polarization at this strain. 

Summarizing, this indicates that, at the same lattice parameter along the polarization direction, the FE properties of SMO and SBMO are nearly identical, if calculated with the same magnetic order.
However, Ba-substitution alters the electronic structure and thus the magnetic coupling, leading to different magnetic order at high strain. Due to the enormous magnetoelectric coupling found in these compounds, this can also drastically affect the electric polarization.

\section{Summary and Conclusions}\label{sec.concl}

We have used DFT+$U$ calculations to study magnetic and ferroelectric properties of Sr$_{1-x}$Ba$_x$MnO$_3$ as functions of isotropic volume expansion and biaxial strain. The calculated magnetic exchange interactions as function of isotropic volume expansion for different Ba concentrations reveal that Ba-substitution has a substantial influence on the electronic structure, which enhances magnetism at a fixed lattice parameter. This is attributed to the stronger hybridization of the more delocalized Ba 5$p$ states with the O $p$ (and $s$) states compared to the  Sr 4$p$ states, which indirectly affects the hybridization between Mn $d$ and O $p$.
This chemical influence is less noticeable on the ferroelectric properties, which instead appear to be determined mostly by the lattice constant in the direction of polarization.

At their equilibrium lattice parameters, both SrMnO$_3$ and Sr$_{0.5}$Ba$_{0.5}$MnO$_3$ exhibit G-type AFM.
However, applying biaxial tensile strain leads to different series of magnetic transitions in the two compounds, again illustrating the difference in magnetic properties caused by Ba-substitution. The ferroelectric polarization appears at similar IP lattice constant for the two compounds and then varies almost identically with strain, as long as the magnetic order remains G-type AFM. However, the strong magnetoelectric coupling causes notably different polarizations as soon as the magnetic orders differ. 

The results of the present study shed new light on the interplay between strain and chemical substitution in the Sr$_{1-x}$Ba$_x$MnO$_3$ system, and the effects on the ferroic properties.
Recent experimental work studied the magnetic ordering temperature, $T_\mathrm{c}$, of  Sr$_{1-x}$Ba$_x$MnO$_3$ thin films and observed that, within the region with G-type AFM, $T_\mathrm{c}$ decreases monotonically with increasing unit cell volume, regardless whether this change in volume is due to strain or chemical substitution. 
Although our calculations reproduce the observed trend in the magnetic ordering temperature with strain or volume, they also indicate that Ba-substitution enhances the magnetic exchange interactions. Hence, at fixed volume, the predicted $T_\mathrm{c}$ is enhanced by Ba substitution within the regime of G-AFM order. 
To resolve this potential disagreement between experiment and theory, further studies, both on the experimental and computational side, are required. Experimentally, it would be desirable to explore more compositions and strain values, ideally within an overlapping volume range. Computationally, the use of 
more sophisticated models describing chemical disorder for $x>0$ that also allow to study intermediate compositions would be instructive.
Furthermore, it would be of interest to investigate anisotropic and antisymmetric exchange interactions and resulting non-collinear magnetic structures, that can be expected in the FE phase with broken inversion symmetry.

\section{Acknowledgments}
This work was supported by the Swiss National Science Foundation (project code 200021E-162297) and the German Science Foundation under the priority program SPP 1599 (``Ferroic Cooling''). Computational work was performed on resources provided by the Swiss National Supercomputing Centre (CSCS) and the ETH Z\"urich.

\bibliography{literature}{}

\begin{thebibliography}{40}%
\makeatletter
\providecommand \@ifxundefined [1]{%
 \@ifx{#1\undefined}
}%
\providecommand \@ifnum [1]{%
 \ifnum #1\expandafter \@firstoftwo
 \else \expandafter \@secondoftwo
 \fi
}%
\providecommand \@ifx [1]{%
 \ifx #1\expandafter \@firstoftwo
 \else \expandafter \@secondoftwo
 \fi
}%
\providecommand \natexlab [1]{#1}%
\providecommand \enquote  [1]{``#1''}%
\providecommand \bibnamefont  [1]{#1}%
\providecommand \bibfnamefont [1]{#1}%
\providecommand \citenamefont [1]{#1}%
\providecommand \href@noop [0]{\@secondoftwo}%
\providecommand \href [0]{\begingroup \@sanitize@url \@href}%
\providecommand \@href[1]{\@@startlink{#1}\@@href}%
\providecommand \@@href[1]{\endgroup#1\@@endlink}%
\providecommand \@sanitize@url [0]{\catcode `\\12\catcode `\$12\catcode
  `\&12\catcode `\#12\catcode `\^12\catcode `\_12\catcode `\%12\relax}%
\providecommand \@@startlink[1]{}%
\providecommand \@@endlink[0]{}%
\providecommand \url  [0]{\begingroup\@sanitize@url \@url }%
\providecommand \@url [1]{\endgroup\@href {#1}{\urlprefix }}%
\providecommand \urlprefix  [0]{URL }%
\providecommand \Eprint [0]{\href }%
\providecommand \doibase [0]{http://dx.doi.org/}%
\providecommand \selectlanguage [0]{\@gobble}%
\providecommand \bibinfo  [0]{\@secondoftwo}%
\providecommand \bibfield  [0]{\@secondoftwo}%
\providecommand \translation [1]{[#1]}%
\providecommand \BibitemOpen [0]{}%
\providecommand \bibitemStop [0]{}%
\providecommand \bibitemNoStop [0]{.\EOS\space}%
\providecommand \EOS [0]{\spacefactor3000\relax}%
\providecommand \BibitemShut  [1]{\csname bibitem#1\endcsname}%
\let\auto@bib@innerbib\@empty
\bibitem [{\citenamefont {Lee}\ and\ \citenamefont
  {Rabe}(2010)}]{PhysRevLett.104.207204}%
  \BibitemOpen
  \bibfield  {author} {\bibinfo {author} {\bibfnamefont {J.~H.}\ \bibnamefont
  {Lee}}\ and\ \bibinfo {author} {\bibfnamefont {K.~M.}\ \bibnamefont {Rabe}},\
  }\href {\doibase 10.1103/PhysRevLett.104.207204} {\bibfield  {journal}
  {\bibinfo  {journal} {Phys. Rev. Lett.}\ }\textbf {\bibinfo {volume} {104}},\
  \bibinfo {pages} {207204} (\bibinfo {year} {2010})}\BibitemShut {NoStop}%
\bibitem [{\citenamefont {Rondinelli}\ \emph {et~al.}(2009)\citenamefont
  {Rondinelli}, \citenamefont {Eidelson},\ and\ \citenamefont
  {Spaldin}}]{PhysRevB.79.205119}%
  \BibitemOpen
  \bibfield  {author} {\bibinfo {author} {\bibfnamefont {J.~M.}\ \bibnamefont
  {Rondinelli}}, \bibinfo {author} {\bibfnamefont {A.~S.}\ \bibnamefont
  {Eidelson}}, \ and\ \bibinfo {author} {\bibfnamefont {N.~A.}\ \bibnamefont
  {Spaldin}},\ }\href {\doibase 10.1103/PhysRevB.79.205119} {\bibfield
  {journal} {\bibinfo  {journal} {Phys. Rev. B}\ }\textbf {\bibinfo {volume}
  {79}},\ \bibinfo {pages} {205119} (\bibinfo {year} {2009})}\BibitemShut
  {NoStop}%
\bibitem [{\citenamefont {Becher}\ \emph {et~al.}(2015)\citenamefont {Becher},
  \citenamefont {Maurel}, \citenamefont {Aschauer}, \citenamefont {Lilienblum},
  \citenamefont {Mag\'{e}}, \citenamefont {Meier}, \citenamefont {Langenberg},
  \citenamefont {Trassin}, \citenamefont {Blasco}, \citenamefont {Krug},
  \citenamefont {Algarabel}, \citenamefont {Spaldin}, \citenamefont {Pardo},\
  and\ \citenamefont {Fiebig}}]{Becher2015}%
  \BibitemOpen
  \bibfield  {author} {\bibinfo {author} {\bibfnamefont {C.}~\bibnamefont
  {Becher}}, \bibinfo {author} {\bibfnamefont {L.}~\bibnamefont {Maurel}},
  \bibinfo {author} {\bibfnamefont {U.}~\bibnamefont {Aschauer}}, \bibinfo
  {author} {\bibfnamefont {M.}~\bibnamefont {Lilienblum}}, \bibinfo {author}
  {\bibfnamefont {C.}~\bibnamefont {Mag\'{e}}}, \bibinfo {author}
  {\bibfnamefont {D.}~\bibnamefont {Meier}}, \bibinfo {author} {\bibfnamefont
  {E.}~\bibnamefont {Langenberg}}, \bibinfo {author} {\bibfnamefont
  {M.}~\bibnamefont {Trassin}}, \bibinfo {author} {\bibfnamefont
  {J.}~\bibnamefont {Blasco}}, \bibinfo {author} {\bibfnamefont {I.~P.}\
  \bibnamefont {Krug}}, \bibinfo {author} {\bibfnamefont {P.~A.}\ \bibnamefont
  {Algarabel}}, \bibinfo {author} {\bibfnamefont {N.~A.}\ \bibnamefont
  {Spaldin}}, \bibinfo {author} {\bibfnamefont {J.~A.}\ \bibnamefont {Pardo}},
  \ and\ \bibinfo {author} {\bibfnamefont {M.}~\bibnamefont {Fiebig}},\ }\href
  {http://dx.doi.org/10.1038/nnano.2015.108} {\bibfield  {journal} {\bibinfo
  {journal} {Nature Nanotechnology}\ }\textbf {\bibinfo {volume} {10}},\
  \bibinfo {pages} {661} (\bibinfo {year} {2015})},\ \bibinfo {note}
  {letter}\BibitemShut {NoStop}%
\bibitem [{\citenamefont {Guzman}\ \emph {et~al.}(2016)\citenamefont {Guzman},
  \citenamefont {Maurel}, \citenamefont {Langenberg}, \citenamefont {Lupini},
  \citenamefont {Algarabel}, \citenamefont {Pardo},\ and\ \citenamefont
  {Magen}}]{acs.nanolett.5b04455}%
  \BibitemOpen
  \bibfield  {author} {\bibinfo {author} {\bibfnamefont {R.}~\bibnamefont
  {Guzman}}, \bibinfo {author} {\bibfnamefont {L.}~\bibnamefont {Maurel}},
  \bibinfo {author} {\bibfnamefont {E.}~\bibnamefont {Langenberg}}, \bibinfo
  {author} {\bibfnamefont {A.~R.}\ \bibnamefont {Lupini}}, \bibinfo {author}
  {\bibfnamefont {P.~A.}\ \bibnamefont {Algarabel}}, \bibinfo {author}
  {\bibfnamefont {J.~A.}\ \bibnamefont {Pardo}}, \ and\ \bibinfo {author}
  {\bibfnamefont {C.}~\bibnamefont {Magen}},\ }\href {\doibase
  10.1021/acs.nanolett.5b04455} {\bibfield  {journal} {\bibinfo  {journal}
  {Nano Letters}\ }\textbf {\bibinfo {volume} {16}},\ \bibinfo {pages} {2221}
  (\bibinfo {year} {2016})}\BibitemShut {NoStop}%
\bibitem [{\citenamefont {Guo}\ \emph {et~al.}(2018)\citenamefont {Guo},
  \citenamefont {Wang}, \citenamefont {Yuan}, \citenamefont {He}, \citenamefont
  {Lu}, \citenamefont {Chen}, \citenamefont {Yang}, \citenamefont {Wang},
  \citenamefont {Erni}, \citenamefont {Rossell}, \citenamefont {Gopalan},
  \citenamefont {Xiang}, \citenamefont {Tokura},\ and\ \citenamefont
  {Yu}}]{PhysRevB.97.235135}%
  \BibitemOpen
  \bibfield  {author} {\bibinfo {author} {\bibfnamefont {J.~W.}\ \bibnamefont
  {Guo}}, \bibinfo {author} {\bibfnamefont {P.~S.}\ \bibnamefont {Wang}},
  \bibinfo {author} {\bibfnamefont {Y.}~\bibnamefont {Yuan}}, \bibinfo {author}
  {\bibfnamefont {Q.}~\bibnamefont {He}}, \bibinfo {author} {\bibfnamefont
  {J.~L.}\ \bibnamefont {Lu}}, \bibinfo {author} {\bibfnamefont {T.~Z.}\
  \bibnamefont {Chen}}, \bibinfo {author} {\bibfnamefont {S.~Z.}\ \bibnamefont
  {Yang}}, \bibinfo {author} {\bibfnamefont {Y.~J.}\ \bibnamefont {Wang}},
  \bibinfo {author} {\bibfnamefont {R.}~\bibnamefont {Erni}}, \bibinfo {author}
  {\bibfnamefont {M.~D.}\ \bibnamefont {Rossell}}, \bibinfo {author}
  {\bibfnamefont {V.}~\bibnamefont {Gopalan}}, \bibinfo {author} {\bibfnamefont
  {H.~J.}\ \bibnamefont {Xiang}}, \bibinfo {author} {\bibfnamefont
  {Y.}~\bibnamefont {Tokura}}, \ and\ \bibinfo {author} {\bibfnamefont
  {P.}~\bibnamefont {Yu}},\ }\href {\doibase 10.1103/PhysRevB.97.235135}
  {\bibfield  {journal} {\bibinfo  {journal} {Phys. Rev. B}\ }\textbf {\bibinfo
  {volume} {97}},\ \bibinfo {pages} {235135} (\bibinfo {year}
  {2018})}\BibitemShut {NoStop}%
\bibitem [{\citenamefont {Sakai}\ \emph {et~al.}(2011)\citenamefont {Sakai},
  \citenamefont {Fujioka}, \citenamefont {Fukuda}, \citenamefont {Okuyama},
  \citenamefont {Hashizume}, \citenamefont {Kagawa}, \citenamefont {Nakao},
  \citenamefont {Murakami}, \citenamefont {Arima}, \citenamefont {Baron},
  \citenamefont {Taguchi},\ and\ \citenamefont
  {Tokura}}]{PhysRevLett.107.137601}%
  \BibitemOpen
  \bibfield  {author} {\bibinfo {author} {\bibfnamefont {H.}~\bibnamefont
  {Sakai}}, \bibinfo {author} {\bibfnamefont {J.}~\bibnamefont {Fujioka}},
  \bibinfo {author} {\bibfnamefont {T.}~\bibnamefont {Fukuda}}, \bibinfo
  {author} {\bibfnamefont {D.}~\bibnamefont {Okuyama}}, \bibinfo {author}
  {\bibfnamefont {D.}~\bibnamefont {Hashizume}}, \bibinfo {author}
  {\bibfnamefont {F.}~\bibnamefont {Kagawa}}, \bibinfo {author} {\bibfnamefont
  {H.}~\bibnamefont {Nakao}}, \bibinfo {author} {\bibfnamefont
  {Y.}~\bibnamefont {Murakami}}, \bibinfo {author} {\bibfnamefont
  {T.}~\bibnamefont {Arima}}, \bibinfo {author} {\bibfnamefont {A.~Q.~R.}\
  \bibnamefont {Baron}}, \bibinfo {author} {\bibfnamefont {Y.}~\bibnamefont
  {Taguchi}}, \ and\ \bibinfo {author} {\bibfnamefont {Y.}~\bibnamefont
  {Tokura}},\ }\href {\doibase 10.1103/PhysRevLett.107.137601} {\bibfield
  {journal} {\bibinfo  {journal} {Phys. Rev. Lett.}\ }\textbf {\bibinfo
  {volume} {107}},\ \bibinfo {pages} {137601} (\bibinfo {year}
  {2011})}\BibitemShut {NoStop}%
\bibitem [{\citenamefont {Lee}\ and\ \citenamefont
  {Rabe}(2011)}]{PhysRevB.84.104440}%
  \BibitemOpen
  \bibfield  {author} {\bibinfo {author} {\bibfnamefont {J.~H.}\ \bibnamefont
  {Lee}}\ and\ \bibinfo {author} {\bibfnamefont {K.~M.}\ \bibnamefont {Rabe}},\
  }\href {\doibase 10.1103/PhysRevB.84.104440} {\bibfield  {journal} {\bibinfo
  {journal} {Phys. Rev. B}\ }\textbf {\bibinfo {volume} {84}},\ \bibinfo
  {pages} {104440} (\bibinfo {year} {2011})}\BibitemShut {NoStop}%
\bibitem [{\citenamefont {Hong}\ \emph {et~al.}(2012)\citenamefont {Hong},
  \citenamefont {Stroppa}, \citenamefont {\'I\~niguez}, \citenamefont
  {Picozzi},\ and\ \citenamefont {Vanderbilt}}]{PhysRevB.85.054417}%
  \BibitemOpen
  \bibfield  {author} {\bibinfo {author} {\bibfnamefont {J.}~\bibnamefont
  {Hong}}, \bibinfo {author} {\bibfnamefont {A.}~\bibnamefont {Stroppa}},
  \bibinfo {author} {\bibfnamefont {J.}~\bibnamefont {\'I\~niguez}}, \bibinfo
  {author} {\bibfnamefont {S.}~\bibnamefont {Picozzi}}, \ and\ \bibinfo
  {author} {\bibfnamefont {D.}~\bibnamefont {Vanderbilt}},\ }\href {\doibase
  10.1103/PhysRevB.85.054417} {\bibfield  {journal} {\bibinfo  {journal} {Phys.
  Rev. B}\ }\textbf {\bibinfo {volume} {85}},\ \bibinfo {pages} {054417}
  (\bibinfo {year} {2012})}\BibitemShut {NoStop}%
\bibitem [{\citenamefont {Giovannetti}\ \emph {et~al.}(2012)\citenamefont
  {Giovannetti}, \citenamefont {Kumar}, \citenamefont {Ortix}, \citenamefont
  {Capone},\ and\ \citenamefont {van~den Brink}}]{PhysRevLett.109.107601}%
  \BibitemOpen
  \bibfield  {author} {\bibinfo {author} {\bibfnamefont {G.}~\bibnamefont
  {Giovannetti}}, \bibinfo {author} {\bibfnamefont {S.}~\bibnamefont {Kumar}},
  \bibinfo {author} {\bibfnamefont {C.}~\bibnamefont {Ortix}}, \bibinfo
  {author} {\bibfnamefont {M.}~\bibnamefont {Capone}}, \ and\ \bibinfo {author}
  {\bibfnamefont {J.}~\bibnamefont {van~den Brink}},\ }\href {\doibase
  10.1103/PhysRevLett.109.107601} {\bibfield  {journal} {\bibinfo  {journal}
  {Phys. Rev. Lett.}\ }\textbf {\bibinfo {volume} {109}},\ \bibinfo {pages}
  {107601} (\bibinfo {year} {2012})}\BibitemShut {NoStop}%
\bibitem [{\citenamefont {Sakai}\ \emph {et~al.}(2012)\citenamefont {Sakai},
  \citenamefont {Fujioka}, \citenamefont {Fukuda}, \citenamefont {Bahramy},
  \citenamefont {Okuyama}, \citenamefont {Arita}, \citenamefont {Arima},
  \citenamefont {Baron}, \citenamefont {Taguchi},\ and\ \citenamefont
  {Tokura}}]{Sakai_et_al:2012}%
  \BibitemOpen
  \bibfield  {author} {\bibinfo {author} {\bibfnamefont {H.}~\bibnamefont
  {Sakai}}, \bibinfo {author} {\bibfnamefont {J.}~\bibnamefont {Fujioka}},
  \bibinfo {author} {\bibfnamefont {T.}~\bibnamefont {Fukuda}}, \bibinfo
  {author} {\bibfnamefont {M.~S.}\ \bibnamefont {Bahramy}}, \bibinfo {author}
  {\bibfnamefont {D.}~\bibnamefont {Okuyama}}, \bibinfo {author} {\bibfnamefont
  {R.}~\bibnamefont {Arita}}, \bibinfo {author} {\bibfnamefont
  {T.}~\bibnamefont {Arima}}, \bibinfo {author} {\bibfnamefont {A.~Q.~R.}\
  \bibnamefont {Baron}}, \bibinfo {author} {\bibfnamefont {Y.}~\bibnamefont
  {Taguchi}}, \ and\ \bibinfo {author} {\bibfnamefont {Y.}~\bibnamefont
  {Tokura}},\ }\href {\doibase 10.1103/physrevb.86.104407} {\bibfield
  {journal} {\bibinfo  {journal} {Physical Review B}\ }\textbf {\bibinfo
  {volume} {86}},\ \bibinfo {pages} {104407} (\bibinfo {year}
  {2012})}\BibitemShut {NoStop}%
\bibitem [{\citenamefont {Kamba}\ \emph {et~al.}(2014)\citenamefont {Kamba},
  \citenamefont {Goian}, \citenamefont {Skoromets}, \citenamefont
  {Hejtm\'anek}, \citenamefont {Bovtun}, \citenamefont {Kempa}, \citenamefont
  {Borodavka}, \citenamefont {Van\ifmmode~\check{e}\else \v{e}\fi{}k},
  \citenamefont {Belik}, \citenamefont {Lee}, \citenamefont {Pacherov\'a},\
  and\ \citenamefont {Rabe}}]{PhysRevB.89.064308}%
  \BibitemOpen
  \bibfield  {author} {\bibinfo {author} {\bibfnamefont {S.}~\bibnamefont
  {Kamba}}, \bibinfo {author} {\bibfnamefont {V.}~\bibnamefont {Goian}},
  \bibinfo {author} {\bibfnamefont {V.}~\bibnamefont {Skoromets}}, \bibinfo
  {author} {\bibfnamefont {J.}~\bibnamefont {Hejtm\'anek}}, \bibinfo {author}
  {\bibfnamefont {V.}~\bibnamefont {Bovtun}}, \bibinfo {author} {\bibfnamefont
  {M.}~\bibnamefont {Kempa}}, \bibinfo {author} {\bibfnamefont
  {F.}~\bibnamefont {Borodavka}}, \bibinfo {author} {\bibfnamefont
  {P.}~\bibnamefont {Van\ifmmode~\check{e}\else \v{e}\fi{}k}}, \bibinfo
  {author} {\bibfnamefont {A.~A.}\ \bibnamefont {Belik}}, \bibinfo {author}
  {\bibfnamefont {J.~H.}\ \bibnamefont {Lee}}, \bibinfo {author} {\bibfnamefont
  {O.}~\bibnamefont {Pacherov\'a}}, \ and\ \bibinfo {author} {\bibfnamefont
  {K.~M.}\ \bibnamefont {Rabe}},\ }\href {\doibase 10.1103/PhysRevB.89.064308}
  {\bibfield  {journal} {\bibinfo  {journal} {Phys. Rev. B}\ }\textbf {\bibinfo
  {volume} {89}},\ \bibinfo {pages} {064308} (\bibinfo {year}
  {2014})}\BibitemShut {NoStop}%
\bibitem [{\citenamefont {Goian}\ \emph {et~al.}(2017)\citenamefont {Goian},
  \citenamefont {Langenberg}, \citenamefont {Marcano}, \citenamefont {Bovtun},
  \citenamefont {Maurel}, \citenamefont {Kempa}, \citenamefont {Prokscha},
  \citenamefont {Kroupa}, \citenamefont {Algarabel}, \citenamefont {Pardo},\
  and\ \citenamefont {Kamba}}]{Goian_et_al:2017}%
  \BibitemOpen
  \bibfield  {author} {\bibinfo {author} {\bibfnamefont {V.}~\bibnamefont
  {Goian}}, \bibinfo {author} {\bibfnamefont {E.}~\bibnamefont {Langenberg}},
  \bibinfo {author} {\bibfnamefont {N.}~\bibnamefont {Marcano}}, \bibinfo
  {author} {\bibfnamefont {V.}~\bibnamefont {Bovtun}}, \bibinfo {author}
  {\bibfnamefont {L.}~\bibnamefont {Maurel}}, \bibinfo {author} {\bibfnamefont
  {M.}~\bibnamefont {Kempa}}, \bibinfo {author} {\bibfnamefont
  {T.}~\bibnamefont {Prokscha}}, \bibinfo {author} {\bibfnamefont
  {J.}~\bibnamefont {Kroupa}}, \bibinfo {author} {\bibfnamefont {P.~A.}\
  \bibnamefont {Algarabel}}, \bibinfo {author} {\bibfnamefont {J.~A.}\
  \bibnamefont {Pardo}}, \ and\ \bibinfo {author} {\bibfnamefont
  {S.}~\bibnamefont {Kamba}},\ }\href {\doibase 10.1103/physrevb.95.075126}
  {\bibfield  {journal} {\bibinfo  {journal} {Physical Review B}\ }\textbf
  {\bibinfo {volume} {95}},\ \bibinfo {pages} {075126} (\bibinfo {year}
  {2017})}\BibitemShut {NoStop}%
\bibitem [{\citenamefont {Edstr\"om}\ and\ \citenamefont
  {Ederer}(2018)}]{PhysRevMaterials.2.104409}%
  \BibitemOpen
  \bibfield  {author} {\bibinfo {author} {\bibfnamefont {A.}~\bibnamefont
  {Edstr\"om}}\ and\ \bibinfo {author} {\bibfnamefont {C.}~\bibnamefont
  {Ederer}},\ }\href {\doibase 10.1103/PhysRevMaterials.2.104409} {\bibfield
  {journal} {\bibinfo  {journal} {Phys. Rev. Materials}\ }\textbf {\bibinfo
  {volume} {2}},\ \bibinfo {pages} {104409} (\bibinfo {year}
  {2018})}\BibitemShut {NoStop}%
\bibitem [{\citenamefont {Edstr\"om}\ and\ \citenamefont
  {Ederer}(2020)}]{2019arXiv190512955E}%
  \BibitemOpen
  \bibfield  {author} {\bibinfo {author} {\bibfnamefont {A.}~\bibnamefont
  {Edstr\"om}}\ and\ \bibinfo {author} {\bibfnamefont {C.}~\bibnamefont
  {Ederer}},\ }\href {\doibase 10.1103/PhysRevLett.124.167201} {\bibfield
  {journal} {\bibinfo  {journal} {Phys. Rev. Lett.}\ }\textbf {\bibinfo
  {volume} {124}},\ \bibinfo {pages} {167201} (\bibinfo {year}
  {2020})}\BibitemShut {NoStop}%
\bibitem [{\citenamefont {Pratt}\ \emph {et~al.}(2014)\citenamefont {Pratt},
  \citenamefont {Lynn}, \citenamefont {Mais}, \citenamefont {Chmaissem},
  \citenamefont {Brown}, \citenamefont {Kolesnik},\ and\ \citenamefont
  {Dabrowski}}]{PhysRevB.90.140401}%
  \BibitemOpen
  \bibfield  {author} {\bibinfo {author} {\bibfnamefont {D.~K.}\ \bibnamefont
  {Pratt}}, \bibinfo {author} {\bibfnamefont {J.~W.}\ \bibnamefont {Lynn}},
  \bibinfo {author} {\bibfnamefont {J.}~\bibnamefont {Mais}}, \bibinfo {author}
  {\bibfnamefont {O.}~\bibnamefont {Chmaissem}}, \bibinfo {author}
  {\bibfnamefont {D.~E.}\ \bibnamefont {Brown}}, \bibinfo {author}
  {\bibfnamefont {S.}~\bibnamefont {Kolesnik}}, \ and\ \bibinfo {author}
  {\bibfnamefont {B.}~\bibnamefont {Dabrowski}},\ }\href {\doibase
  10.1103/PhysRevB.90.140401} {\bibfield  {journal} {\bibinfo  {journal} {Phys.
  Rev. B}\ }\textbf {\bibinfo {volume} {90}},\ \bibinfo {pages} {140401}
  (\bibinfo {year} {2014})}\BibitemShut {NoStop}%
\bibitem [{\citenamefont {Langenberg}\ \emph {et~al.}(2015)\citenamefont
  {Langenberg}, \citenamefont {Guzm{\'{a}}n}, \citenamefont {Maurel},
  \citenamefont {Mart\'{i}nez~de Ba{\~{n}}os}, \citenamefont {Morell{\'{o}}n},
  \citenamefont {Ibarra}, \citenamefont {Herrero-Mart\'{i}n}, \citenamefont
  {Blasco}, \citenamefont {Mag{\'{e}}n}, \citenamefont {Algarabel},\ and\
  \citenamefont {Pardo}}]{doi:10.1021/acsami.5b06478}%
  \BibitemOpen
  \bibfield  {author} {\bibinfo {author} {\bibfnamefont {E.}~\bibnamefont
  {Langenberg}}, \bibinfo {author} {\bibfnamefont {R.}~\bibnamefont
  {Guzm{\'{a}}n}}, \bibinfo {author} {\bibfnamefont {L.}~\bibnamefont
  {Maurel}}, \bibinfo {author} {\bibfnamefont {L.}~\bibnamefont
  {Mart\'{i}nez~de Ba{\~{n}}os}}, \bibinfo {author} {\bibfnamefont
  {L.}~\bibnamefont {Morell{\'{o}}n}}, \bibinfo {author} {\bibfnamefont
  {M.~R.}\ \bibnamefont {Ibarra}}, \bibinfo {author} {\bibfnamefont
  {J.}~\bibnamefont {Herrero-Mart\'{i}n}}, \bibinfo {author} {\bibfnamefont
  {J.}~\bibnamefont {Blasco}}, \bibinfo {author} {\bibfnamefont
  {C.}~\bibnamefont {Mag{\'{e}}n}}, \bibinfo {author} {\bibfnamefont {P.~A.}\
  \bibnamefont {Algarabel}}, \ and\ \bibinfo {author} {\bibfnamefont {J.~A.}\
  \bibnamefont {Pardo}},\ }\href {\doibase 10.1021/acsami.5b06478} {\bibfield
  {journal} {\bibinfo  {journal} {ACS Applied Materials \& Interfaces}\
  }\textbf {\bibinfo {volume} {7}},\ \bibinfo {pages} {23967} (\bibinfo {year}
  {2015})},\ \bibinfo {note} {pMID: 26462710}\BibitemShut {NoStop}%
\bibitem [{\citenamefont {Langenberg}\ \emph {et~al.}(2017)\citenamefont
  {Langenberg}, \citenamefont {Maurel}, \citenamefont {Marcano}, \citenamefont
  {Guzman}, \citenamefont {Strichovanec}, \citenamefont {Prokscha},
  \citenamefont {Magen}, \citenamefont {Algarabel},\ and\ \citenamefont
  {Pardo}}]{ADMI201601040}%
  \BibitemOpen
  \bibfield  {author} {\bibinfo {author} {\bibfnamefont {E.}~\bibnamefont
  {Langenberg}}, \bibinfo {author} {\bibfnamefont {L.}~\bibnamefont {Maurel}},
  \bibinfo {author} {\bibfnamefont {N.}~\bibnamefont {Marcano}}, \bibinfo
  {author} {\bibfnamefont {R.}~\bibnamefont {Guzman}}, \bibinfo {author}
  {\bibfnamefont {P.}~\bibnamefont {Strichovanec}}, \bibinfo {author}
  {\bibfnamefont {T.}~\bibnamefont {Prokscha}}, \bibinfo {author}
  {\bibfnamefont {C.}~\bibnamefont {Magen}}, \bibinfo {author} {\bibfnamefont
  {P.~A.}\ \bibnamefont {Algarabel}}, \ and\ \bibinfo {author} {\bibfnamefont
  {J.~A.}\ \bibnamefont {Pardo}},\ }\href {\doibase 10.1002/admi.201601040}
  {\bibfield  {journal} {\bibinfo  {journal} {Advanced Materials Interfaces}\
  }\textbf {\bibinfo {volume} {4}},\ \bibinfo {pages} {1601040} (\bibinfo
  {year} {2017})},\ \bibinfo {note} {1601040}\BibitemShut {NoStop}%
\bibitem [{\citenamefont {Maurel}\ \emph {et~al.}(2019)\citenamefont {Maurel},
  \citenamefont {Marcano}, \citenamefont {Langenberg}, \citenamefont
  {Guzm{\'{a}}n}, \citenamefont {Prokscha}, \citenamefont {Mag{\'{e}}n},
  \citenamefont {Pardo},\ and\ \citenamefont
  {Algarabel}}]{doi:10.1063/1.5090824}%
  \BibitemOpen
  \bibfield  {author} {\bibinfo {author} {\bibfnamefont {L.}~\bibnamefont
  {Maurel}}, \bibinfo {author} {\bibfnamefont {N.}~\bibnamefont {Marcano}},
  \bibinfo {author} {\bibfnamefont {E.}~\bibnamefont {Langenberg}}, \bibinfo
  {author} {\bibfnamefont {R.}~\bibnamefont {Guzm{\'{a}}n}}, \bibinfo {author}
  {\bibfnamefont {T.}~\bibnamefont {Prokscha}}, \bibinfo {author}
  {\bibfnamefont {C.}~\bibnamefont {Mag{\'{e}}n}}, \bibinfo {author}
  {\bibfnamefont {J.~A.}\ \bibnamefont {Pardo}}, \ and\ \bibinfo {author}
  {\bibfnamefont {P.~A.}\ \bibnamefont {Algarabel}},\ }\href {\doibase
  10.1063/1.5090824} {\bibfield  {journal} {\bibinfo  {journal} {APL
  Materials}\ }\textbf {\bibinfo {volume} {7}},\ \bibinfo {pages} {041117}
  (\bibinfo {year} {2019})},\ \Eprint
  {http://arxiv.org/abs/https://doi.org/10.1063/1.5090824}
  {https://doi.org/10.1063/1.5090824} \BibitemShut {NoStop}%
\bibitem [{\citenamefont {Nourafkan}\ \emph {et~al.}(2014)\citenamefont
  {Nourafkan}, \citenamefont {Kotliar},\ and\ \citenamefont
  {Tremblay}}]{PhysRevB.90.220405}%
  \BibitemOpen
  \bibfield  {author} {\bibinfo {author} {\bibfnamefont {R.}~\bibnamefont
  {Nourafkan}}, \bibinfo {author} {\bibfnamefont {G.}~\bibnamefont {Kotliar}},
  \ and\ \bibinfo {author} {\bibfnamefont {A.-M.~S.}\ \bibnamefont
  {Tremblay}},\ }\href {\doibase 10.1103/PhysRevB.90.220405} {\bibfield
  {journal} {\bibinfo  {journal} {Phys. Rev. B}\ }\textbf {\bibinfo {volume}
  {90}},\ \bibinfo {pages} {220405} (\bibinfo {year} {2014})}\BibitemShut
  {NoStop}%
\bibitem [{\citenamefont {Chen}\ and\ \citenamefont
  {Millis}(2016{\natexlab{a}})}]{PhysRevB.94.165106}%
  \BibitemOpen
  \bibfield  {author} {\bibinfo {author} {\bibfnamefont {H.}~\bibnamefont
  {Chen}}\ and\ \bibinfo {author} {\bibfnamefont {A.~J.}\ \bibnamefont
  {Millis}},\ }\href {\doibase 10.1103/PhysRevB.94.165106} {\bibfield
  {journal} {\bibinfo  {journal} {Phys. Rev. B}\ }\textbf {\bibinfo {volume}
  {94}},\ \bibinfo {pages} {165106} (\bibinfo {year}
  {2016}{\natexlab{a}})}\BibitemShut {NoStop}%
\bibitem [{\citenamefont {Bayaraa}\ \emph {et~al.}(2018)\citenamefont
  {Bayaraa}, \citenamefont {Yang}, \citenamefont {Zhao}, \citenamefont
  {\'I\~niguez},\ and\ \citenamefont {Bellaiche}}]{PhysRevMaterials.2.084404}%
  \BibitemOpen
  \bibfield  {author} {\bibinfo {author} {\bibfnamefont {T.}~\bibnamefont
  {Bayaraa}}, \bibinfo {author} {\bibfnamefont {Y.}~\bibnamefont {Yang}},
  \bibinfo {author} {\bibfnamefont {H.~J.}\ \bibnamefont {Zhao}}, \bibinfo
  {author} {\bibfnamefont {J.}~\bibnamefont {\'I\~niguez}}, \ and\ \bibinfo
  {author} {\bibfnamefont {L.}~\bibnamefont {Bellaiche}},\ }\href {\doibase
  10.1103/PhysRevMaterials.2.084404} {\bibfield  {journal} {\bibinfo  {journal}
  {Phys. Rev. Materials}\ }\textbf {\bibinfo {volume} {2}},\ \bibinfo {pages}
  {084404} (\bibinfo {year} {2018})}\BibitemShut {NoStop}%
\bibitem [{\citenamefont {Bl\"ochl}(1994)}]{PhysRevB.50.17953}%
  \BibitemOpen
  \bibfield  {author} {\bibinfo {author} {\bibfnamefont {P.~E.}\ \bibnamefont
  {Bl\"ochl}},\ }\href {\doibase 10.1103/PhysRevB.50.17953} {\bibfield
  {journal} {\bibinfo  {journal} {Phys. Rev. B}\ }\textbf {\bibinfo {volume}
  {50}},\ \bibinfo {pages} {17953} (\bibinfo {year} {1994})}\BibitemShut
  {NoStop}%
\bibitem [{\citenamefont {Kresse}\ and\ \citenamefont
  {Joubert}(1999)}]{PhysRevB.59.1758}%
  \BibitemOpen
  \bibfield  {author} {\bibinfo {author} {\bibfnamefont {G.}~\bibnamefont
  {Kresse}}\ and\ \bibinfo {author} {\bibfnamefont {D.}~\bibnamefont
  {Joubert}},\ }\href {\doibase 10.1103/PhysRevB.59.1758} {\bibfield  {journal}
  {\bibinfo  {journal} {Phys. Rev. B}\ }\textbf {\bibinfo {volume} {59}},\
  \bibinfo {pages} {1758} (\bibinfo {year} {1999})}\BibitemShut {NoStop}%
\bibitem [{\citenamefont {Kresse}\ and\ \citenamefont
  {Furthm\"uller}(1996)}]{KRESSE199615}%
  \BibitemOpen
  \bibfield  {author} {\bibinfo {author} {\bibfnamefont {G.}~\bibnamefont
  {Kresse}}\ and\ \bibinfo {author} {\bibfnamefont {J.}~\bibnamefont
  {Furthm\"uller}},\ }\href {\doibase
  https://doi.org/10.1016/0927-0256(96)00008-0} {\bibfield  {journal} {\bibinfo
   {journal} {Computational Materials Science}\ }\textbf {\bibinfo {volume}
  {6}},\ \bibinfo {pages} {15 } (\bibinfo {year} {1996})}\BibitemShut {NoStop}%
\bibitem [{\citenamefont {Kresse}\ and\ \citenamefont
  {Hafner}(1994)}]{PhysRevB.49.14251}%
  \BibitemOpen
  \bibfield  {author} {\bibinfo {author} {\bibfnamefont {G.}~\bibnamefont
  {Kresse}}\ and\ \bibinfo {author} {\bibfnamefont {J.}~\bibnamefont
  {Hafner}},\ }\href {\doibase 10.1103/PhysRevB.49.14251} {\bibfield  {journal}
  {\bibinfo  {journal} {Phys. Rev. B}\ }\textbf {\bibinfo {volume} {49}},\
  \bibinfo {pages} {14251} (\bibinfo {year} {1994})}\BibitemShut {NoStop}%
\bibitem [{\citenamefont {Kresse}\ and\ \citenamefont
  {Hafner}(1993)}]{PhysRevB.47.558}%
  \BibitemOpen
  \bibfield  {author} {\bibinfo {author} {\bibfnamefont {G.}~\bibnamefont
  {Kresse}}\ and\ \bibinfo {author} {\bibfnamefont {J.}~\bibnamefont
  {Hafner}},\ }\href {\doibase 10.1103/PhysRevB.47.558} {\bibfield  {journal}
  {\bibinfo  {journal} {Phys. Rev. B}\ }\textbf {\bibinfo {volume} {47}},\
  \bibinfo {pages} {558} (\bibinfo {year} {1993})}\BibitemShut {NoStop}%
\bibitem [{\citenamefont {Perdew}\ \emph {et~al.}(2008)\citenamefont {Perdew},
  \citenamefont {Ruzsinszky}, \citenamefont {Csonka}, \citenamefont {Vydrov},
  \citenamefont {Scuseria}, \citenamefont {Constantin}, \citenamefont {Zhou},\
  and\ \citenamefont {Burke}}]{PhysRevLett.100.136406}%
  \BibitemOpen
  \bibfield  {author} {\bibinfo {author} {\bibfnamefont {J.~P.}\ \bibnamefont
  {Perdew}}, \bibinfo {author} {\bibfnamefont {A.}~\bibnamefont {Ruzsinszky}},
  \bibinfo {author} {\bibfnamefont {G.~I.}\ \bibnamefont {Csonka}}, \bibinfo
  {author} {\bibfnamefont {O.~A.}\ \bibnamefont {Vydrov}}, \bibinfo {author}
  {\bibfnamefont {G.~E.}\ \bibnamefont {Scuseria}}, \bibinfo {author}
  {\bibfnamefont {L.~A.}\ \bibnamefont {Constantin}}, \bibinfo {author}
  {\bibfnamefont {X.}~\bibnamefont {Zhou}}, \ and\ \bibinfo {author}
  {\bibfnamefont {K.}~\bibnamefont {Burke}},\ }\href {\doibase
  10.1103/PhysRevLett.100.136406} {\bibfield  {journal} {\bibinfo  {journal}
  {Phys. Rev. Lett.}\ }\textbf {\bibinfo {volume} {100}},\ \bibinfo {pages}
  {136406} (\bibinfo {year} {2008})}\BibitemShut {NoStop}%
\bibitem [{\citenamefont {Dudarev}\ \emph {et~al.}(1998)\citenamefont
  {Dudarev}, \citenamefont {Botton}, \citenamefont {Savrasov}, \citenamefont
  {Humphreys},\ and\ \citenamefont {Sutton}}]{PhysRevB.57.1505}%
  \BibitemOpen
  \bibfield  {author} {\bibinfo {author} {\bibfnamefont {S.~L.}\ \bibnamefont
  {Dudarev}}, \bibinfo {author} {\bibfnamefont {G.~A.}\ \bibnamefont {Botton}},
  \bibinfo {author} {\bibfnamefont {S.~Y.}\ \bibnamefont {Savrasov}}, \bibinfo
  {author} {\bibfnamefont {C.~J.}\ \bibnamefont {Humphreys}}, \ and\ \bibinfo
  {author} {\bibfnamefont {A.~P.}\ \bibnamefont {Sutton}},\ }\href {\doibase
  10.1103/PhysRevB.57.1505} {\bibfield  {journal} {\bibinfo  {journal} {Phys.
  Rev. B}\ }\textbf {\bibinfo {volume} {57}},\ \bibinfo {pages} {1505}
  (\bibinfo {year} {1998})}\BibitemShut {NoStop}%
\bibitem [{\citenamefont {Marthinsen}\ \emph {et~al.}(2016)\citenamefont
  {Marthinsen}, \citenamefont {Faber}, \citenamefont {Aschauer}, \citenamefont
  {Spaldin},\ and\ \citenamefont {Selbach}}]{marthinsen_2016}%
  \BibitemOpen
  \bibfield  {author} {\bibinfo {author} {\bibfnamefont {A.}~\bibnamefont
  {Marthinsen}}, \bibinfo {author} {\bibfnamefont {C.}~\bibnamefont {Faber}},
  \bibinfo {author} {\bibfnamefont {U.}~\bibnamefont {Aschauer}}, \bibinfo
  {author} {\bibfnamefont {N.~A.}\ \bibnamefont {Spaldin}}, \ and\ \bibinfo
  {author} {\bibfnamefont {S.~M.}\ \bibnamefont {Selbach}},\ }\href {\doibase
  10.1557/mrc.2016.30} {\bibfield  {journal} {\bibinfo  {journal} {MRS
  Communications}\ }\textbf {\bibinfo {volume} {6}},\ \bibinfo {pages} {182}
  (\bibinfo {year} {2016})}\BibitemShut {NoStop}%
\bibitem [{\citenamefont {Zhu}\ \emph {et~al.}(2020)\citenamefont {Zhu},
  \citenamefont {Edstr\"om},\ and\ \citenamefont {Ederer}}]{SMO_Jij}%
  \BibitemOpen
  \bibfield  {author} {\bibinfo {author} {\bibfnamefont {X.}~\bibnamefont
  {Zhu}}, \bibinfo {author} {\bibfnamefont {A.}~\bibnamefont {Edstr\"om}}, \
  and\ \bibinfo {author} {\bibfnamefont {C.}~\bibnamefont {Ederer}},\ }\href
  {\doibase 10.1103/PhysRevB.101.064401} {\bibfield  {journal} {\bibinfo
  {journal} {Phys. Rev. B}\ }\textbf {\bibinfo {volume} {101}},\ \bibinfo
  {pages} {064401} (\bibinfo {year} {2020})}\BibitemShut {NoStop}%
\bibitem [{\citenamefont {Chen}\ and\ \citenamefont
  {Millis}(2016{\natexlab{b}})}]{PhysRevB.93.205110}%
  \BibitemOpen
  \bibfield  {author} {\bibinfo {author} {\bibfnamefont {H.}~\bibnamefont
  {Chen}}\ and\ \bibinfo {author} {\bibfnamefont {A.~J.}\ \bibnamefont
  {Millis}},\ }\href {\doibase 10.1103/PhysRevB.93.205110} {\bibfield
  {journal} {\bibinfo  {journal} {Phys. Rev. B}\ }\textbf {\bibinfo {volume}
  {93}},\ \bibinfo {pages} {205110} (\bibinfo {year}
  {2016}{\natexlab{b}})}\BibitemShut {NoStop}%
\bibitem [{\citenamefont {Perdew}\ \emph {et~al.}(1996)\citenamefont {Perdew},
  \citenamefont {Burke},\ and\ \citenamefont
  {Ernzerhof}}]{PhysRevLett.77.3865}%
  \BibitemOpen
  \bibfield  {author} {\bibinfo {author} {\bibfnamefont {J.~P.}\ \bibnamefont
  {Perdew}}, \bibinfo {author} {\bibfnamefont {K.}~\bibnamefont {Burke}}, \
  and\ \bibinfo {author} {\bibfnamefont {M.}~\bibnamefont {Ernzerhof}},\ }\href
  {\doibase 10.1103/PhysRevLett.77.3865} {\bibfield  {journal} {\bibinfo
  {journal} {Phys. Rev. Lett.}\ }\textbf {\bibinfo {volume} {77}},\ \bibinfo
  {pages} {3865} (\bibinfo {year} {1996})}\BibitemShut {NoStop}%
\bibitem [{\citenamefont {Xiang}\ \emph {et~al.}(2011)\citenamefont {Xiang},
  \citenamefont {Kan}, \citenamefont {Wei}, \citenamefont {Whangbo},\ and\
  \citenamefont {Gong}}]{PhysRevB.84.224429}%
  \BibitemOpen
  \bibfield  {author} {\bibinfo {author} {\bibfnamefont {H.~J.}\ \bibnamefont
  {Xiang}}, \bibinfo {author} {\bibfnamefont {E.~J.}\ \bibnamefont {Kan}},
  \bibinfo {author} {\bibfnamefont {S.-H.}\ \bibnamefont {Wei}}, \bibinfo
  {author} {\bibfnamefont {M.-H.}\ \bibnamefont {Whangbo}}, \ and\ \bibinfo
  {author} {\bibfnamefont {X.~G.}\ \bibnamefont {Gong}},\ }\href {\doibase
  10.1103/PhysRevB.84.224429} {\bibfield  {journal} {\bibinfo  {journal} {Phys.
  Rev. B}\ }\textbf {\bibinfo {volume} {84}},\ \bibinfo {pages} {224429}
  (\bibinfo {year} {2011})}\BibitemShut {NoStop}%
\bibitem [{\citenamefont {King-Smith}\ and\ \citenamefont
  {Vanderbilt}(1993)}]{PhysRevB.47.1651}%
  \BibitemOpen
  \bibfield  {author} {\bibinfo {author} {\bibfnamefont {R.~D.}\ \bibnamefont
  {King-Smith}}\ and\ \bibinfo {author} {\bibfnamefont {D.}~\bibnamefont
  {Vanderbilt}},\ }\href {\doibase 10.1103/PhysRevB.47.1651} {\bibfield
  {journal} {\bibinfo  {journal} {Phys. Rev. B}\ }\textbf {\bibinfo {volume}
  {47}},\ \bibinfo {pages} {1651} (\bibinfo {year} {1993})}\BibitemShut
  {NoStop}%
\bibitem [{Note1()}]{Note1}%
  \BibitemOpen
  \bibinfo {note} {As mentioned in Sec.~\ref {sec.intro}, the synthesis of
  Sr$_{1-x}$Ba$_{x}$MnO$_3$~ in the cubic perovskite structure is only possible
  up to $x \approx 0.5$. We nevertheless include the case of
  perovskite-structured BMO here, analogous to Ref.~\protect \rev@citealpnum
  {PhysRevB.79.205119}, to allow for a more systematic comparison.}\BibitemShut
  {Stop}%
\bibitem [{\citenamefont {Wollan}\ and\ \citenamefont
  {Koehler}(1955)}]{Wollan/Koehler:1955}%
  \BibitemOpen
  \bibfield  {author} {\bibinfo {author} {\bibfnamefont {E.~O.}\ \bibnamefont
  {Wollan}}\ and\ \bibinfo {author} {\bibfnamefont {W.~C.}\ \bibnamefont
  {Koehler}},\ }\href@noop {} {\bibfield  {journal} {\bibinfo  {journal}
  {Physical Review}\ }\textbf {\bibinfo {volume} {100}},\ \bibinfo {pages}
  {545} (\bibinfo {year} {1955})}\BibitemShut {NoStop}%
\bibitem [{\citenamefont {Marzari}\ \emph {et~al.}(2012)\citenamefont
  {Marzari}, \citenamefont {Mostofi}, \citenamefont {Yates}, \citenamefont
  {Souza},\ and\ \citenamefont {Vanderbilt}}]{Marzari_et_al:2012}%
  \BibitemOpen
  \bibfield  {author} {\bibinfo {author} {\bibfnamefont {N.}~\bibnamefont
  {Marzari}}, \bibinfo {author} {\bibfnamefont {A.~A.}\ \bibnamefont
  {Mostofi}}, \bibinfo {author} {\bibfnamefont {J.~R.}\ \bibnamefont {Yates}},
  \bibinfo {author} {\bibfnamefont {I.}~\bibnamefont {Souza}}, \ and\ \bibinfo
  {author} {\bibfnamefont {D.}~\bibnamefont {Vanderbilt}},\ }\href {\doibase
  10.1103/RevModPhys.84.1419} {\bibfield  {journal} {\bibinfo  {journal}
  {Reviews of Modern Physics}\ }\textbf {\bibinfo {volume} {84}},\ \bibinfo
  {pages} {1419} (\bibinfo {year} {2012})}\BibitemShut {NoStop}%
\bibitem [{\citenamefont {Maurel}\ \emph {et~al.}(2015)\citenamefont {Maurel},
  \citenamefont {Marcano}, \citenamefont {Prokscha}, \citenamefont
  {Langenberg}, \citenamefont {Blasco}, \citenamefont {Guzm\'an}, \citenamefont
  {Suter}, \citenamefont {Mag\'en}, \citenamefont {Morell\'on}, \citenamefont
  {Ibarra}, \citenamefont {Pardo},\ and\ \citenamefont
  {Algarabel}}]{PhysRevB.92.024419}%
  \BibitemOpen
  \bibfield  {author} {\bibinfo {author} {\bibfnamefont {L.}~\bibnamefont
  {Maurel}}, \bibinfo {author} {\bibfnamefont {N.}~\bibnamefont {Marcano}},
  \bibinfo {author} {\bibfnamefont {T.}~\bibnamefont {Prokscha}}, \bibinfo
  {author} {\bibfnamefont {E.}~\bibnamefont {Langenberg}}, \bibinfo {author}
  {\bibfnamefont {J.}~\bibnamefont {Blasco}}, \bibinfo {author} {\bibfnamefont
  {R.}~\bibnamefont {Guzm\'an}}, \bibinfo {author} {\bibfnamefont
  {A.}~\bibnamefont {Suter}}, \bibinfo {author} {\bibfnamefont
  {C.}~\bibnamefont {Mag\'en}}, \bibinfo {author} {\bibfnamefont
  {L.}~\bibnamefont {Morell\'on}}, \bibinfo {author} {\bibfnamefont {M.~R.}\
  \bibnamefont {Ibarra}}, \bibinfo {author} {\bibfnamefont {J.~A.}\
  \bibnamefont {Pardo}}, \ and\ \bibinfo {author} {\bibfnamefont {P.~A.}\
  \bibnamefont {Algarabel}},\ }\href {\doibase 10.1103/PhysRevB.92.024419}
  {\bibfield  {journal} {\bibinfo  {journal} {Phys. Rev. B}\ }\textbf {\bibinfo
  {volume} {92}},\ \bibinfo {pages} {024419} (\bibinfo {year}
  {2015})}\BibitemShut {NoStop}%
\bibitem [{\citenamefont {Okuyama}\ \emph {et~al.}(2020)\citenamefont
  {Okuyama}, \citenamefont {Yamauchi}, \citenamefont {Sakai}, \citenamefont
  {Taguchi}, \citenamefont {Tokura}, \citenamefont {Sugimoto}, \citenamefont
  {Sato},\ and\ \citenamefont {Oguchi}}]{okuyama2020}%
  \BibitemOpen
  \bibfield  {author} {\bibinfo {author} {\bibfnamefont {D.}~\bibnamefont
  {Okuyama}}, \bibinfo {author} {\bibfnamefont {K.}~\bibnamefont {Yamauchi}},
  \bibinfo {author} {\bibfnamefont {H.}~\bibnamefont {Sakai}}, \bibinfo
  {author} {\bibfnamefont {Y.}~\bibnamefont {Taguchi}}, \bibinfo {author}
  {\bibfnamefont {Y.}~\bibnamefont {Tokura}}, \bibinfo {author} {\bibfnamefont
  {K.}~\bibnamefont {Sugimoto}}, \bibinfo {author} {\bibfnamefont {T.~J.}\
  \bibnamefont {Sato}}, \ and\ \bibinfo {author} {\bibfnamefont
  {T.}~\bibnamefont {Oguchi}},\ }\href@noop {} {\enquote {\bibinfo {title}
  {Ferroelectric atomic displacement in multiferroic tetragonal perovskite
  sr$_{1/2}$ba$_{1/2}$mno$_3$},}\ } (\bibinfo {year} {2020}),\ \Eprint
  {http://arxiv.org/abs/2002.02198} {arXiv:2002.02198 [cond-mat.mtrl-sci]}
  \BibitemShut {NoStop}%
\bibitem [{hig()}]{higherLambdaNote}%
  \BibitemOpen
  \href@noop {} {}\bibinfo {note} {This is consistent with the results in
  Ref.~\onlinecite{2019arXiv190512955E}, where the biquadratic magnetoelectric
  coupling coefficient for G-AFM order couples more favorably to $P$ than the
  one for C-AFM, at 4\% strain. However, a direct quantitative comparison is
  not possible, since the biquadratic coupling was found to be insufficient at
  4\% strain and low temperature, where higher order terms are
  important.}\BibitemShut {Stop}%
\end{thebibliography}%
\bibliographystyle{apsrev4-1}

\end{document}